\documentclass[aps,prd,twocolumn,
               notitlepage,mathrsfs,
               amssymb,amsmath,amsfonts,
               superscriptaddress,
               longbibliography,
               nofootinbib,floatfix]{revtex4-2}

\usepackage{graphicx}
\usepackage{url}
\usepackage[linktocpage,breaklinks]{hyperref}
\usepackage[capitalize]{cleveref}
\usepackage[usenames,dvipsnames]{xcolor}
\hypersetup{colorlinks=true,
            citecolor=NavyBlue,
            linkcolor=magenta,
            urlcolor=magenta}
\usepackage{multirow,array}
\DeclareMathAlphabet{\pazocal}{OMS}{zplm}{m}{n}
\usepackage{journals}

\begin{document}

\title{Gravitational waves from core collapse of rotating very-massive stars: \\
3D numerical relativity computation}

\date{\today}

\author{Alan Tsz-Lok Lam}
\email{tpl5641@psu.edu}
\affiliation{Institute for Gravitation and the Cosmos, The Pennsylvania State University, University Park, PA 16802, USA}
\affiliation{Department of Physics, The Pennsylvania State University, University Park, PA 16802, USA}
\affiliation{Max-Planck-Institut f\"ur Gravitationsphysik (Albert-Einstein-Institut), Am M\"uhlenberg 1, D-14476 Potsdam-Golm, Germany}

\author{Masaru Shibata}
\affiliation{Max-Planck-Institut f\"ur Gravitationsphysik (Albert-Einstein-Institut), Am M\"uhlenberg 1, D-14476 Potsdam-Golm, Germany}
\affiliation{Center of Gravitational Physics and Quantum Information, Yukawa Institute for Theoretical Physics, Kyoto University, Kyoto, 606-8502, Japan} 

\author{Sho Fujibayashi}
\affiliation{Frontier Research Institute for Interdisciplinary Sciences, Tohoku University, Sendai 980-8578, Japan}
\affiliation{Astronomical Institute, Tohoku University, Aoba, Sendai 980-8578, Japan}
\affiliation{Max-Planck-Institut f\"ur Gravitationsphysik (Albert-Einstein-Institut), Am M\"uhlenberg 1, D-14476 Potsdam-Golm, Germany}

\begin{abstract}
We numerically study the collapse of rotating, very-massive stellar cores with masses of $\approx 200$, $300$, $500$, and $1100M_\odot$ into black holes using both axisymmetric and three-dimensional (3D) numerical relativity. Our results indicate that when the dimensionless spin of the resulting black hole exceeds 0.8, a massive disk consistently forms around it. These massive, compact disks, carrying more than about 10\% of the black hole's mass, are prone to non-axisymmetric deformations that trigger gravitational-wave bursts with frequencies around 10-50 Hz. Such waves could be detected by the Einstein Telescope and Cosmic Explorer, even from sources a few Gpc away. We also summarize the gravitational-wave signals from axisymmetric collapse, which tend to have lower amplitudes and higher frequencies than those caused by non-axisymmetric instabilities. 
\end{abstract}

\maketitle

\section{Introduction}\label{sec1}

Very-massive stars with initial masses larger than $100M_\odot$ are believed to form at the high-mass end of the stars in low-metalicity environments~(see, e.g., Refs.~\cite{Hirano2014feb, Susa2014sep, Hirano2015mar} for an early theoretical attempt and \cite{2026arXiv260627427P} for the latest status). In particular, the latest simulation study~\cite{2026arXiv260627427P} indicates that the typical mass accretion rate in the central region of proto-galaxies is $\sim 10^{-2}M_\odot/\mathrm{yr}$. If this accretion persists for the lifetime of massive stars ($\sim 10^6$\,yrs) and an appreciable mass fraction falls into the central region, a very-massive star with $\sim 10^3$--$10^4M_\odot$ could form. Metal-poor very-massive stars with initial masses larger than $\sim 260M_\odot$ and less than $10^4M_\odot$ are believed to collapse into a black hole after the onset of the electron-positron pair-creation instability (pair instability)~\cite{1971reas.book.....Z, 1996snih.book.....A, 2001ApJ...550..372F, Heger:2001cd} (while for more massive stars with masses larger than $\sim 10^4M_\odot$, i.e., supermassive stars, collapse occurs via a general-relativistic instability~\cite{1964ApJ...140..417C}). Very-massive stars evolve through hydrogen and helium burning, eventually forming a core composed primarily of oxygen \cite{Bond:1984sn, 1996snih.book.....A, 2012A&A...542A.113Y, Takahashi2018}. The hot, heavy oxygen core finally becomes unstable to the pair instability, leading to a black hole, as illustrated by fully general-relativistic simulations~\cite{Uchida:2018ago, Shibata:2025lde}. This is one plausible scenario for the formation of intermediate-mass black holes and seed black holes of supermassive black holes in the early universe~\cite{Inayoshi2020aug, 2020ARA&A..58..257G}. 

In the presence of sufficient angular momentum, the collapse of very-massive stars could serve as an engine for multi-messenger signals because a disk should form around the newly formed black hole~\cite{Uchida:2018ago, Shibata:2025lde}. If a massive disk exists, the remnant might produce energetic events such as gamma-ray bursts and powerful supernovae~(e.g., Refs.~\cite{2001ApJ...550..372F, Heger:2001cd, Uchida:2018ago, Siegel:2021ptt, 2025arXiv250815887G}). Such a system can also be a strong gravitational-wave emitter if the remnant disk is massive enough, i.e., the disk mass is larger than $\sim 10\%$ of the black hole mass~\cite{2011PhRvD..83d3007K, Kiuchi:2011re, Shibata:2021sau}, because massive disks are unstable to non-axisymmetric deformation (see also Refs.~\cite{1984MNRAS.208..721P, 1986PThPh..75..251K, 1991ApJ...381..496H}). The formation process of such rapidly rotating, very-massive stars remains uncertain (see, e.g., Refs.~\cite{Tanikawa:2025fxw, Croon:2025gol, Stegmann:2025cja, Popa:2025dpz, Kiroglu:2025vqy}). Achieving very rapid rotation through standard massive star formation via mass accretion is unlikely (e.g., Ref.~\cite{2023ApJ...950..184K}) because of the so-called $\Omega\Gamma$-limit~\cite{Lee_2016}. However, they are probably formed at least at the centers of dense star clusters through runaway stellar mergers~\cite{2004Natur.428..724P, Tanikawa:2026kcd}. 

Reference~\cite{LIGOScientific:2025rsn} reported the discovery of a high-mass binary black hole (GW231123) by gravitational-wave observatories, with estimated masses of $137^{+22}_{-17}M_\odot$ and $103^{+20}_{-52}M_\odot$ for the two black holes. The estimated dimensionless spins for each black hole are quite high, at $0.90^{+0.10}_{-0.19}$ and $0.80^{+0.20}_{-0.51}$, with 90\% credible intervals. This finding suggests the existence of rapidly spinning intermediate-mass black holes. Near the pair instability threshold, the core mass of very-massive stars is about half their initial stellar mass (e.g., Refs.~\cite{2016MNRAS.456.1320T, Takahashi2018}). Furthermore, the collapse of rapidly rotating progenitors can leave a disk around the black hole, as we demonstrate, which can lead to disk winds that eject significant matter and reduce the black hole's final mass. Therefore, the black holes associated with GW231123, with masses around 100--$150M_\odot$, may have formed from the collapse of a rapidly rotating, very-massive stellar core with an initial mass $\agt 260M_\odot$, a likely requirement for stellar collapse rather than explosion~\cite{Takahashi2018}.

This paper aims to systematically examine the processes of the core collapse of rotating very-massive stars and to show that collapses of rapidly rotating very-massive stellar cores with mass $\sim 200$--$10^3M_\odot$ can indeed form a massive disk around the resulting black hole if the core's surface rotates at an angular velocity $\agt 30\%$ of the Keplerian value. Furthermore, we demonstrate that the remnant massive disk is unstable to non-axisymmetric deformations and is a strong gravitational-wave emitter. We also show that such gravitational waves can be promising sources for the Einstein Telescope (ET)~\citep{Hild:2010id} and the Cosmic Explorer (CE)~\cite{Reitze:2019iox} even if the distance to the source is $\sim 1$\,Gpc or more. As a first step toward more detailed studies, we employ a simplified equation of state that still captures the essential processes of very-massive stellar core collapse. We do not consider low-frequency gravitational waves below 1\,Hz because they lie outside the ET and CE bands, although such waves can be emitted during the free-fall motion of the core collapse and by neutrino emission as a memory effect~\cite{2007ApJ...665L..43S}. 

Throughout this paper, we use the geometric units $c=1=G$, where $c$ and $G$ are the speed of light and the gravitational constant, respectively. $k$ denotes the Boltzmann constant. To clarify the units, we sometimes restore $c$ and $G$. 

\section{Initial data: Stellar core models}\label{sec2}

\begin{figure}
    \centering
    \includegraphics[width=0.99\columnwidth]{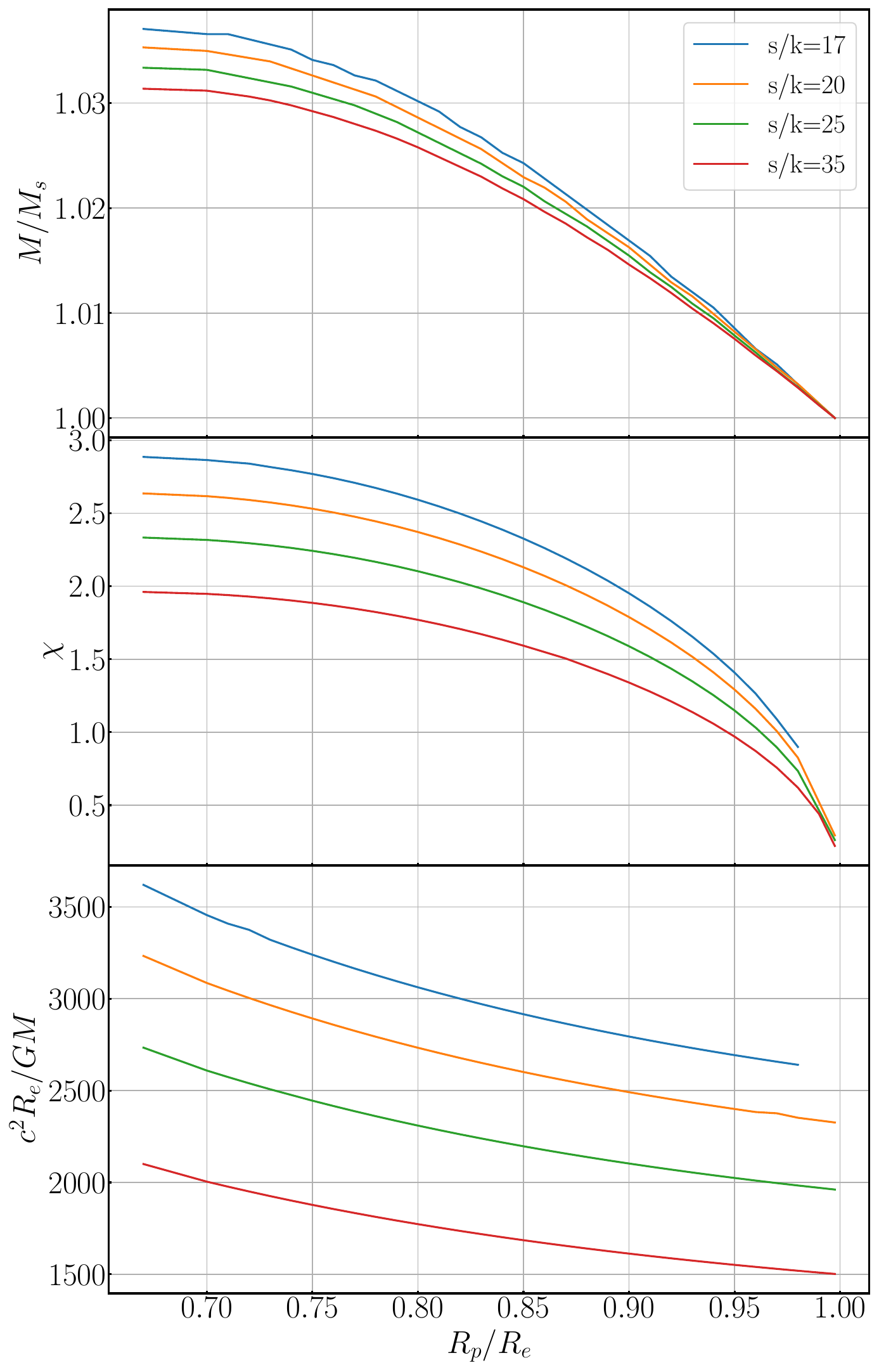}
    \caption{Mass (top panel), dimensionless spin (middle panel), and equatorial radius in units of $GM/c^2(=r_\mathrm{g})$ (bottom panel) as functions of axial ratio, $R_\mathrm{p}/R_\mathrm{e}$, for the stellar core models with $s/k=17$, 20, 25, and 35. For the top panel, $M_\mathrm{s}/M_\odot\approx 203$, 299, 503, and 1074, for $s/k=17$, 20, 25, and 35, respectively.
    }
    \label{fig0}
\end{figure}

Following Ref.~\cite{Shibata:2025lde}, we prepare marginally stable, rotating stellar cores composed of fully ionized oxygen, electrons, and photons in equilibrium as initial conditions for the numerical simulation. We assume rigid rotation for the stellar core models. We employ a Timmes equation of state~\cite{timmes2000a} with a constant entropy per baryon $s$. In this work, we select models with $s/k=17$, 20, 25, and 35 (referred to as S17, S20, S25, and S35, respectively). 
For each case, the approximate mass of the oxygen core is $M\approx 200$, 300, 500, and $1100M_\odot$. 

Figure~\ref{fig0} shows the mass, dimensionless spin, and equatorial radius in units of $r_\mathrm{g}=GM/c^2$, as functions of the axial ratio, $R_\mathrm{p}/R_\mathrm{e}$, for stellar core models with $s/k=17$, 20, 25, and 35. Here, $R_\mathrm{p}$ and $R_\mathrm{e}$ denote the polar and equatorial stellar radii, respectively. Stellar cores with $R_\mathrm{p}/R_\mathrm{e}\approx 2/3$ are approximately located at the mass-shedding limit (i.e., the equatorial stellar surface has nearly Keplerian rotational velocity), implying the allowed range $2/3 \alt R_\mathrm{p}/R_\mathrm{e} \leq 1$. For the top panel, we define the mass of the non-rotating equilibrium core for a given entropy, $M_\mathrm{s}$, as $M_\mathrm{s}/M_\odot\approx 203$, 299, 503, and 1074 for $s/k=17$, 20, 25, and 35, respectively. The mass of the stellar core depends only weakly on the degree of rotation for a given $s/k$; for rotating cases, the mass is larger, but the difference between the non-rotating and maximally rotating cases is within 4\%. The dimensionless spin of the stellar cores is defined by $\chi=cJ/GM^2$, where $J$ is the angular momentum. This can far exceed unity when $R_\mathrm{p}/R_\mathrm{e}$ is smaller than 0.9. This is a noteworthy result given the large stellar core radii at the onset of stellar core collapse~\cite{Shibata:2025lde}. Note that for $R_\mathrm{p}/R_\mathrm{e}=0.95$ and 0.88, the angular velocity at the surface of the stellar cores is approximately 30\% and 50\% of the Keplerian rotational velocity, respectively, irrespective of $s/k$~\cite{Shibata:2025lde}. In this paper, we typically focus on the non-axisymmetric instability of disks for models with $R_\mathrm{p}/R_\mathrm{e}=0.90 \pm 0.05$ (see Sec.~\ref{sec4.4}). 

For a given value of $R_\mathrm{p}/R_\mathrm{e}$, the dimensionless spin is higher in lower-mass models. This is because the stellar core radius in units of $r_\mathrm{g}$ is larger for lower-mass models (see the bottom panel of Fig.~\ref{fig0}). As analyzed in Ref.~\cite{Shibata:2025lde}, for models with $\chi \agt 1$, the black holes formed after stellar core collapse are likely to have a dimensionless spin exceeding 0.8 (see also Sec.~\ref{sec5}); this indicates the formation of rapidly rotating black holes. Additionally, the matter outside the black hole accounts for over 10\% of $M$, suggesting the formation of a massive disk. Consequently, both rapidly and moderately rotating stellar cores are potential progenitors for the formation of rapidly spinning black holes and massive disks.  

\section{Equations of state}\label{sec3}

\begin{figure}
    \centering
    \includegraphics[width=0.95\columnwidth]{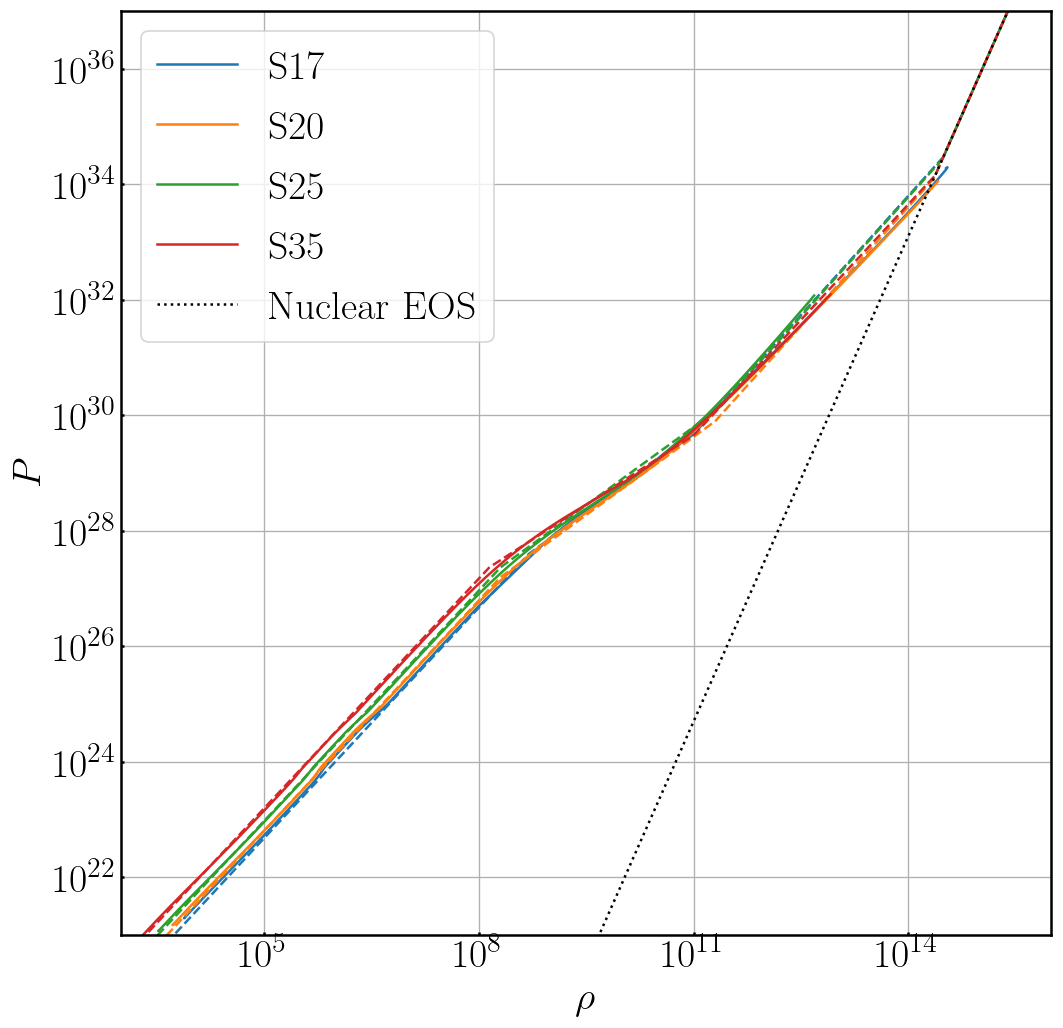}
    \caption{Model equations of state in the plane of $\rho$ and $P$ in the cgs units. The solid and dashed lines show the results from test simulations and the constructed model equations of state, respectively. The dotted line is the model nuclear equation of state for $\rho \agt 2\times 10^{14}\,\mathrm{g/cm^3}$. 
    }
    \label{fig_eos}
\end{figure}

After the onset of pair instability, the core of very-massive stars begins to collapse. In the early stage, the collapse proceeds adiabatically because dissipation processes such as neutrino emission play essentially no role. During this stage, the central temperature and pressure increase along an equation of state with a constant value of $s$. However, at high densities $\agt 10^8\,\mathrm{g/cm^3}$, cooling by neutrino emission becomes important. This cooling significantly suppresses the rise in temperature and pressure. At even higher densities $\agt 10^{10\text{--11}}\,\mathrm{g/cm^3}$, neutrinos become trapped, i.e., their diffusion timescale exceeds the dynamical timescale. Then the collapse again proceeds approximately adiabatically. Because of the high temperature, photo-dissociation also occurs during the collapse, increasing gas pressure~\cite{Sekiguchi2011, Uchida:2018ago}. This also contributes to stiffening the equation of state. 

To account for all these processes, we need to perform a neutrino-radiation hydrodynamics simulation that includes changes in the matter composition. However, multidimensional simulations of stellar core collapse up to black hole formation for a variety of parameters are computationally expensive if we rely solely on radiation hydrodynamics simulations in general relativity. Thus, as a first step, we employ a simplified approach in this paper to qualitatively capture the essence of the core collapse processes. A more detailed quantitative study that directly accounts for all the relevant physics will be conducted in future work. 

More specifically, in the present work, we first perform an axisymmetric radiation hydrodynamics simulation using the code developed by Fujibayashi~\cite{Fujibayashi:2024vnb, Shibata:2025lde, Fujibayashi:2026}. In this code, a Timmes equation of state~\cite{timmes2000a} is implemented, incorporating neutrino cooling, which significantly affects the dynamics of the collapse when the central density is $\rho\sim 10^8$--$10^{11}\,\mathrm{g/cm^3}$. For these simulations, the effect of varying elemental composition is also taken into account. We perform simulations for slowly rotating cases for each value of $s/k$ using this code. From these simulations, we obtain the evolution of the central density and central pressure for each stellar core model (see Fig.~\ref{fig_eos}). Then, we derive fitting formulae for the relation between the rest-mass density $\rho$ and pressure $P$ using the piecewise-polytropic prescription,
\begin{equation}
P=K_i \rho^{\Gamma_i},~~~i=1,2,\cdots,
\end{equation}
where $K_i$ and $\Gamma_i$ denote the polytropic constants and adiabatic indexes for each segment. Here, the pressure must be continuous at the density boundaries of each segment, which are denoted by $\rho_i$ with $i=1, 2,\cdots$, where $\rho_i$ denotes the maximum density of each segment $i$, i.e., $K_i \rho_i^{\Gamma_i}=K_{i+1}\rho_i^{\Gamma_{i+1}}$. Thus, for given values of $K_1$ and $\rho_i$, the equation of state is determined. 

For relatively low-mass very-massive stars, the central density can exceed the nuclear saturation density $\sim 2.8\times 10^{14}\,\mathrm{g/cm^3}$ before a black hole forms. This region ultimately collapses into a black hole on a dynamical timescale, so the nuclear equation of state does not significantly influence the collapse dynamics of very-massive stars. Additionally, the precise equation of state for high-density nuclear matter remains uncertain. Therefore, in this paper, we phenomenologically approximate the high-density region with a polytropic equation of state, $P \propto \rho^{2.8}$. 

In this paper, we employ piecewise polytropic equations of state with 4 pieces. With 4 pieces, the collapse processes described above are captured with good accuracy, at least qualitatively. The equations of state depend on the mass (initial entropy per baryon of the collapsing core). The parameters are summarized in cgs units as follows: For $s/k=17$ (S17 models), 
\begin{align}
    K_1 &= 1.0814 \times 10^{16}, \\
    \Gamma_i &= [ 1.330, 0.9, 1.375, 2.8],\\
    \rho_i &= [5\rho_8, \rho_{11}, 3.0967610519638455\rho_{14}],
\end{align}
for $s/k=20$ (S20 models),
\begin{align}
    K_1 &= 1.46675 \times 10^{16}, \\
    \Gamma_i &= [1.329, 0.9, 1.375, 2.8], \\
    \rho_i &= [2\rho_8, 2\rho_{11}, 2.27778418567055\rho_{14}], 
\end{align}
for $s/k=25$ (S25 models),
\begin{align}
    K_1 &= 2.17140 \times 10^{16}, \\
    \Gamma_i &= [1.329, 0.9, 1.375, 2.8], \\
    \rho_i &= [2\rho_8, 2\rho_{11}, 2.9999741419149\rho_{14}]. 
\end{align}
and $s/k=35$ (S35 models), 
\begin{align}
    K_1 &= 3.76953 \times 10^{16}, \\
    \Gamma_i &= [1.325, 0.8, 1.333, 2.8], \\
    \rho_i &= [1.4\rho_8, \rho_{11}, 2.38216602157632\rho_{14}],
\end{align}
where $\rho_8=10^8\,\mathrm{g/cm^3}$, $\rho_{11}=10^{11}\,\mathrm{g/cm^3}$, and $\rho_{14}=10^{14}\,\mathrm{g/cm^3}$. For $\rho_i$ and $\Gamma_i$, the values are described for $i=1$--3 and for $i=1$--4, respectively. The first segment is a good approximation for the corresponding Timmes equation of state with a constant value of $s$, and the low $\Gamma_2$ reflects the significant neutrino cooling in this density range. Additionally, we verified that even in rotating collapses, the equations of state remain largely consistent with these expressions. 

We perform axisymmetric simulations across a wide range of rotation using these four $s/k$ cases.
We note that our baseline simulations with neutrino cooling were performed using a simple leakage scheme~\cite{Fujibayashi:2026}. The details of neutrino cooling may modify the values of $\Gamma_3$ and $\rho_2$~\cite{2021MNRAS.508..828N}. However, a detailed study of neutrino radiation transfer in multidimensional simulations is beyond the scope of this paper.

In the numerical simulation, we employ a hybrid equation of state of the form
\begin{equation}
P=P_\mathrm{cold}(\rho) + (\Gamma_\mathrm{th} -1) \rho [\varepsilon - \varepsilon_\mathrm{cold}(\rho)], \label{eq11}
\end{equation}
where $P_\mathrm{cold}$ denotes the piecewise polytropic equation of state described above, and $\varepsilon_\mathrm{cold}$ is determined from the first law of thermodynamics, $d\varepsilon_\mathrm{cold}=-P_\mathrm{cold}d \rho^{-1}$, assuming that $\varepsilon_\mathrm{cold}$ is continuous at $\rho=\rho_i$. $\Gamma_\mathrm{th}$ is simply set to $4/3$.
We note that the second term in Eq.~\eqref{eq11} contributes only when shocks form. In the present problem, this occurs only during the formation of the disk around the black hole and of spheroids (see Sec.~\ref{sec4} for the definition of the spheroids). 

\section{Numerical results}\label{sec4}

\subsection{Brief summary of numerical methods}

In this study, we conduct both axisymmetric and non-axisymmetric (3D) simulations using {\tt SACRA-Collapse}, an extension of the original axisymmetric {\tt SACRA-2D}~\cite{Lam:2025pmz}, which is developed for fully general relativistic simulations of a variety of stellar collapses. This extension incorporates (1D) spherical symmetry (refer to S.1. of \cite{Kuan:2026duo}) and 3D Cartesian coordinates, with the same two-to-one fixed-mesh refinement as {\tt SACRA-2D}.

In the original version of {\tt SACRA-2D}, we implemented a fixed-mesh refinement algorithm. However, for very-massive stellar core collapses, this is not computationally suitable because the mass of the black holes at their formation is much smaller than the total mass of the system, $M$, and hence, an extremely high resolution with the finest grid spacing of $\sim 10^{-4}r_\mathrm{g}$ is necessary for some of models, in particular for rapidly rotating models to resolve the black hole while covering the outer part of the stellar core with radius $2000$--$3000\,r_\mathrm{g}$. To overcome this problem, we implement an adaptive scheme in which the finer-resolution zones are added as the central density increases. 
For the simulations presented in this paper, we initially prepare 6 fixed refinement levels covering at least twice the stellar radius.
As the collapse proceeds, we add one refinement level at the center whenever the maximum density increases by a factor of 8.
The metric quantities are prolonged using 6th-order Lagrange interpolation, while the hydrodynamical conserved quantities are prolonged with min-mod limiter following \cite{2020ApJS..249....4S} to conserve total fluid mass and momentum.
At the formation of a black hole, the typical number of refinement levels rises to 17--20, depending on the black hole mass at its formation.
To track the growth of the black hole at formation, refinement levels will be added if the apparent horizon radius is less than 0.25 times the finest domain size.
Because the final black hole might increase in size by as much as a hundred times from its initial formation, we also remove the finest domain whenever the apparent horizon's radius exceeds 0.8 times the size of the domain size. This approach allows us to efficiently track the black hole's growth over a long period.
With this new algorithm, it becomes feasible to resolve the newly formed black hole even if the black hole mass is $10^{-2}M$ with reasonable computational costs (see Sec.~\ref{sec4.2}). 

Using the new {\tt SACRA-Collapse}, we first examine how the collapse proceeds until the formation (or non-formation) of a black hole and how the collapse scenarios depend on the degree of initial rotation and the core mass by performing axisymmetric simulations. 
Then we perform 3D numerical simulations, again using {\tt SACRA-Collapse}, for selected models in which a black hole is surrounded by a massive disk, in a Cartesian-grid ($x, y, z$) setting with fixed mesh refinement optimized for the stellar collapse problem. 

The 3D simulations are performed using numerical results from the axisymmetric simulations carried out in advance as the initial data, following an early work~\cite{Shibata2005}. Specifically, we export the results obtained by the axisymmetric simulations to 3D Cartesian coordinates at an early stage of black hole growth, when the massive disk has not yet fully developed. The initial condition for the 3D simulation is obtained via 6th-order Lagrange interpolation from the axisymmetric data for metric quantities, and linear interpolation for matter.
Although the initial data is axisymmetric, the 3D simulation is performed on Cartesian grids; hence, a non-axisymmetric random perturbation is automatically incorporated. 

In {\tt SACRA-Collapse}, we employ a two-to-one fixed-mesh-refinement structure in the computational domain, composed of a hierarchy of nested concentric grids that overlay one another. The computational domain of axisymmetric simulations covers the region of $[0:x_{\rm max}]$ and $[0:z_{\rm max}]$ for $x$ and $z$, respectively, where $x$ denotes the cylindrical coordinate. $x_\mathrm{max}=z_\mathrm{max}$ is assumed. 
Each adaptive mesh domain contains an even number of grids $N$ in the $x$ and $z$ directions, with the grid spacing written as
\begin{align}
\begin{split}
    \Delta x^{(0)}=\Delta z^{(0)} &= x_{\max} / N, \\
    \Delta x^{(l)}=\Delta z^{(l)} &= \Delta x^{(l-1)} / 2, 
\end{split}
\end{align}
for $l=1,2,\cdots,L-1$. Levels $0$ and $(L-1)$ represent the coarsest and finest levels, respectively. As we already mentioned, in the present study, $L$ varies depending on the situation in the collapse and black hole growth. $N$ is chosen to be 256. At the formation of black holes, the equatorial coordinate radius of the apparent horizon is typically covered by $\sim 15$ grid points in the finest domain. We immediately increase the 2 finer domains so that the black hole horizon is covered by $\sim 60$ grid points in the finest domain, with which the formation and subsequent evolution of the small-mass black holes can be accurately computed.

In the 3D simulation by {\tt SACRA-Collapse}, the computational domain in Cartesian coordinates is $[-x_{\rm max}:x_{\rm max}]$, $[-y_\mathrm{max}:y_\mathrm{max}]$, and $[0:z_{\rm max}]$ where $x_\mathrm{max}=y_\mathrm{max}=z_\mathrm{max}$, which are about 60\% of $x_\mathrm{max}=z_\mathrm{max}$ of the corresponding axisymmetric simulations (see Table~\ref{tab_resolution}). 
The number of refinement levels is chosen so that the equatorial radius of the black hole horizons is resolved by more than 16 grid points in the finest domain at the start of 3D simulations. Since the black hole mass steeply increases in an early stage of the 3D simulations (cf, Fig.~\ref{fig_bh_evo}), the accuracy for resolving the black hole is subsequently improved in the present fixed mesh refinement scheme.

All simulations in this paper assume symmetry with respect to the equatorial plane. Hydrodynamic and geometric variables are defined at cell centers. The HLLC Riemann solver \cite{Lam:2025pmz,Kiuchi:2022ubj,Mignone:2005ft} and the piecewise-parabolic reconstruction scheme \cite{Colella:1982ee} are employed for the general-relativistic hydrodynamics, while the Z4c scheme \cite{Bernuzzi:2009ex} with the moving puncture gauge condition \cite{Alcubierre:2002kk,Baker:2005vv,Campanelli:2005dd} for the metric evolution. For the moving puncture gauge, the damping parameter $\eta_B \sim 2/M$ in the shift condition should be proportional to the inverse of the total mass of the system, which can become stiff at the coarse level with adaptive time-stepping. To manage the stiff damping term in the shift condition, we used the IMEX42L(4,4,2) scheme \cite{Izquierdo:2022eaz}, developed in Ref.~\cite{Lam:2026fcs}.

\begin{table}
\caption{List of grid settings for each model. The columns are the computational domain of axisymmetric simulations, $x_{\max}^{\rm 2D}$, and 3D Cartesian simulation $x_{\max}^{\rm 3D}$, and their corresponding initial finest grid sizes, $\Delta x_{\rm fine, ini}^{\rm 2D}$ and $\Delta x_{\rm fine, ini}^{\rm 3D}$.
\label{tab_resolution}
}
\begin{tabular}{p{0.12\linewidth}|p{0.18\linewidth}p{0.18\linewidth}|p{0.18\linewidth}p{0.18\linewidth}p{0.18\linewidth}|}
\hline
Model &
$x_{\max}^{\rm 2D}$ $(10^6\,{\rm km})$ &
$\Delta x_{\rm fine, ini}^{\rm 2D}$ $(10^2\,{\rm km})$ &
$x_{\max}^{\rm 3D}$ $(10^6\,{\rm km})$ &
$\Delta x_{\rm fine, ini}^{\rm 3D}$ $(10^{-2}\,{\rm km})$ \\ \hline
S17   & 1.77 & 2.16 & 1.11 & 1.65 \\ \hline
S20   & 2.95 & 3.60 & 1.77 & 2.64 \\ \hline
S25   & 2.95 & 3.60 & 1.77 & 2.64 \\ \hline
S35   & 5.91 & 7.21 & 3.69 & 2.75 \\ \hline 
\end{tabular}
\end{table}

\begin{figure*}
    \centering
    \includegraphics[width=\textwidth]{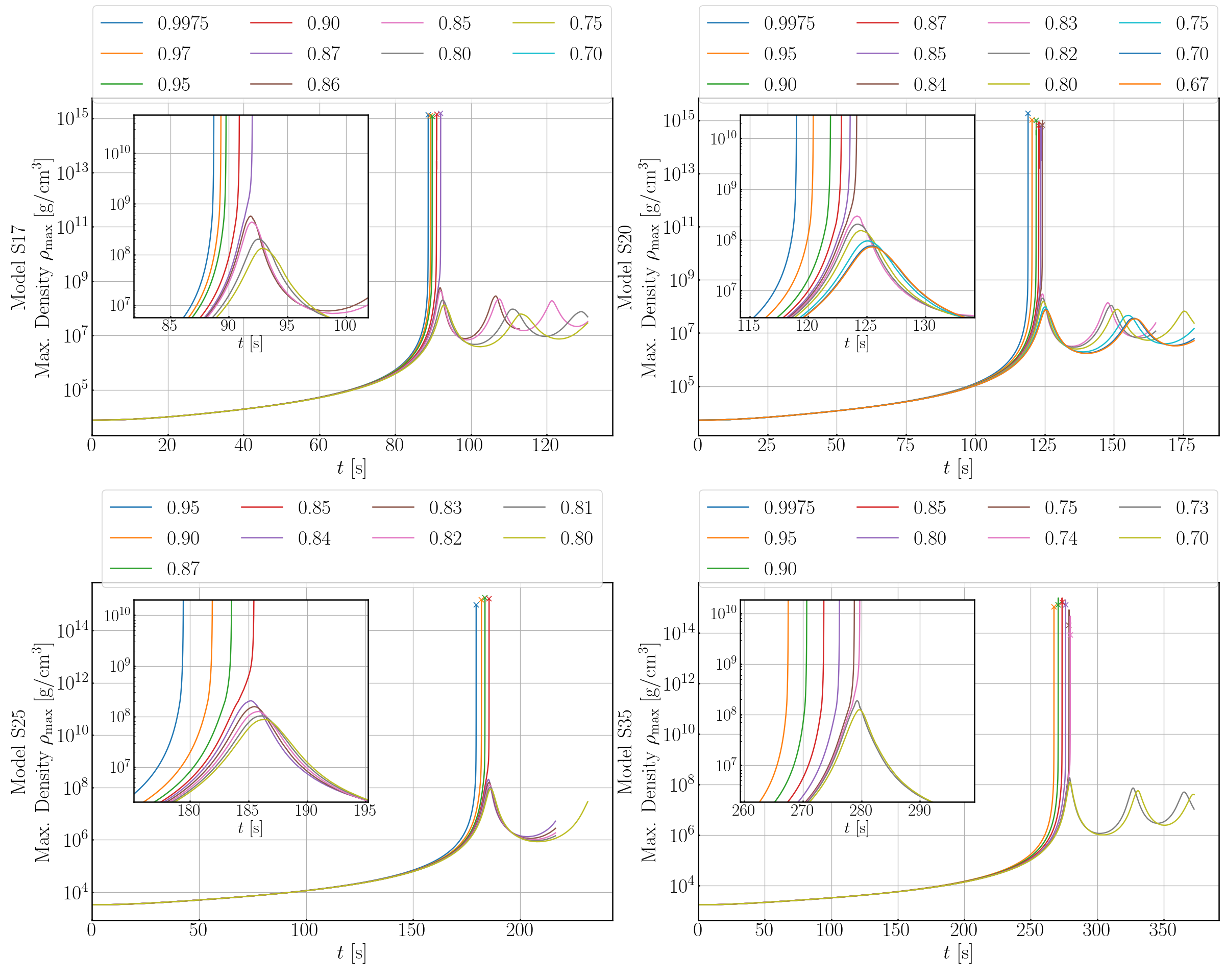}
    \caption{From top left to bottom right, evolution of the central density $\rho_\mathrm{max}$ for models S17, S20, S25 and S35 with a variety of $R_\mathrm{p}/R_\mathrm{e}$. Note that for models with $(s/k, R_\mathrm{p}/R_\mathrm{e})=(17, 0.87)$, $(20, 0.84)$, $(25, 0.85)$, and $(35, 0,74)$, the density rise is decelerated for a short period at $\rho_\mathrm{max}\alt 10^9\,\mathrm{g/cm^3}$, reflecting a weak bounce. 
    The inset in the top-left corner of each plot shows the zoom-in for the evolution of $\rho_\mathrm{max}=10^7$--$10^{10}\,\mathrm{g/cm^{3}}$.
    }
    \label{fig_rhomax_2d}
\end{figure*}

\begin{figure*}
    \centering
    \includegraphics[width=\textwidth]{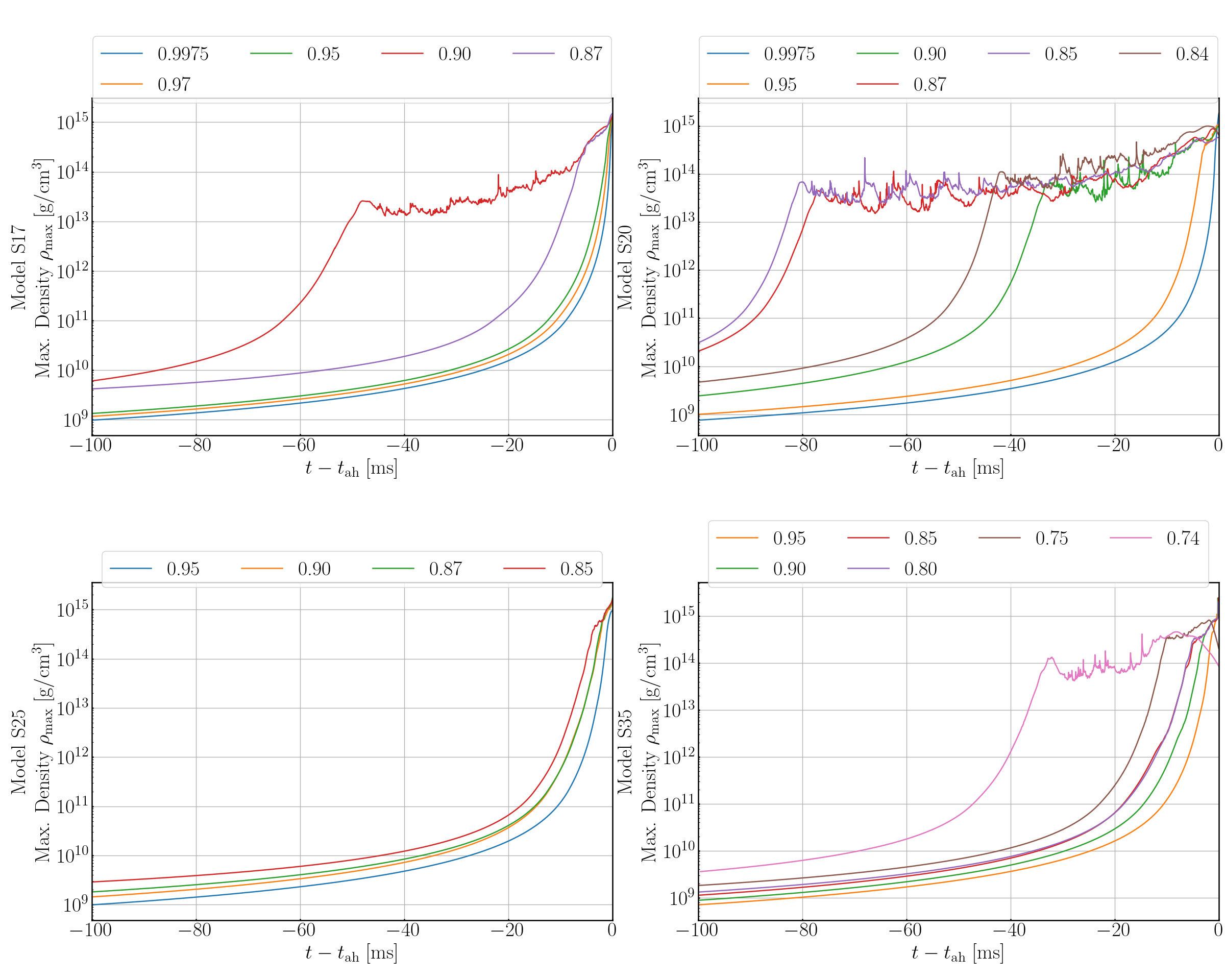}
    \caption{Evolution of the maximum density for 100\,ms before the black hole formation for models S17, S20, S25, and S35 with a variety of $R_\mathrm{p}/R_\mathrm{e}$.
    }
    \label{fig_rhomax_ah}
\end{figure*}

\subsection{Overall evolution processes in the collapse}\label{sec4.2}

\begin{figure*}
    \centering
    \includegraphics[width=\textwidth]{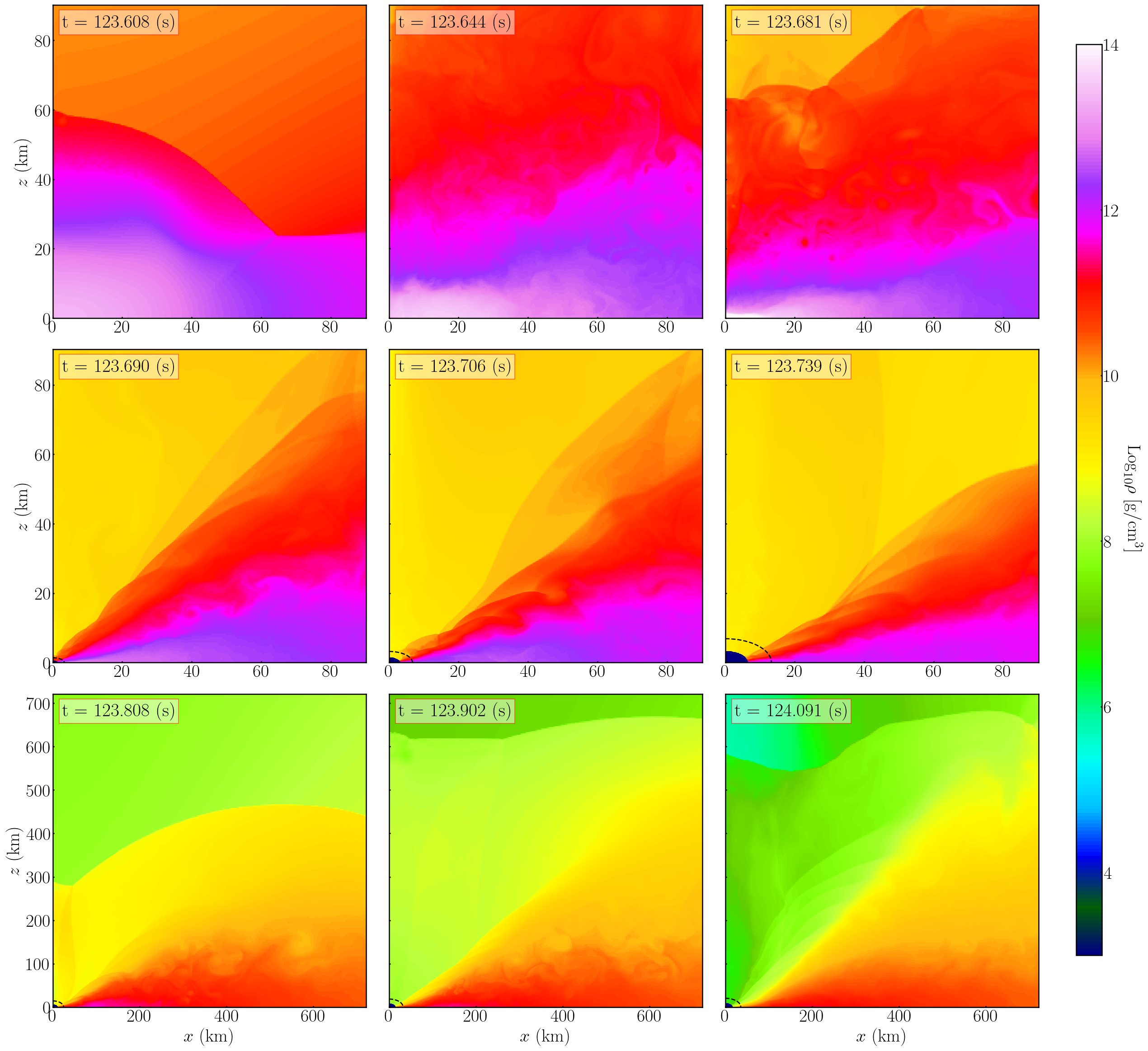}
    \caption{Snapshots of the density profile in the $x$-$z$ plane for model S20 with $R_\mathrm{p}/R_\mathrm{e}=0.85$ near the formation time of a black hole. Note that the scale for the last three panels differs from that of the first six panels. The filled region at the center denotes a computationally excised region, and the dashed circles denote apparent horizons. The density is shown in the unit of $\mathrm{g/cm^3}$. Animations for this model are found at \url{https://www2.yukawa.kyoto-u.ac.jp/~masaru.shibata/movierho20.085a.mp4} (for the inner region) and \url{https://www2.yukawa.kyoto-u.ac.jp/~masaru.shibata/movierho20.085b.mp4} (for a larger region).
    }
    \label{fig_2dcont}
\end{figure*}

We first summarize the findings from the axisymmetric simulations, focusing primarily on the process leading up to black hole formation.
Broadly speaking, there are two classes of outcomes in the very-massive stellar core collapse studied in this paper (see Fig.~\ref{fig_rhomax_2d} for the evolution of the central density in the axisymmetric simulations)\footnote{For relatively low-mass very-massive stars, the collapse may be halted by nuclear burning, leading to a pair-unstable explosion, as mentioned in Sec. I. This can occur for $s/k \alt 15$ (Fujibayashi et al., in preparation). We do not address such cases in this paper.}: (i) A black hole is formed. This is the case for non-rapidly rotating stellar core models; (ii) No black hole forms, but an oscillating spheroid is formed with a maximum density $\rho_\mathrm{max}\alt 10^9\,\mathrm{g/cm^3}$. This occurs only in rapidly rotating stellar core models in which the centrifugal bounce occurs before the equation of state becomes very soft due to neutrino cooling. 

The class (ii) collapse tends to occur in relatively low-mass stellar core models, which can sustain higher degrees of stellar rotation due to their large initial stellar core radii. For example, for model S17, the class (ii) collapse occurs for $R_\mathrm{p}/R_\mathrm{e} \leq 0.86$, whereas for model S35 it occurs only in rapidly rotating models with $R_\mathrm{p}/R_\mathrm{e} \leq 0.73$ (note that the minimum value of $R_\mathrm{p}/R_\mathrm{e}$ is $\approx 2/3$ for the rigidly rotating progenitor core models). In this paper, we focus only on the class (i) in the 3D simulations because the spheroid formed in the class (ii) has a maximum density of $10^7$--$10^9\,\mathrm{g/cm^3}$, and for such an object, neutrino cooling and nuclear reactions, which are only phenomenologically treated in the present study, can play a crucial role in the subsequent long-term evolution.

\begin{figure*}
    \centering
    \includegraphics[width=0.49\textwidth]{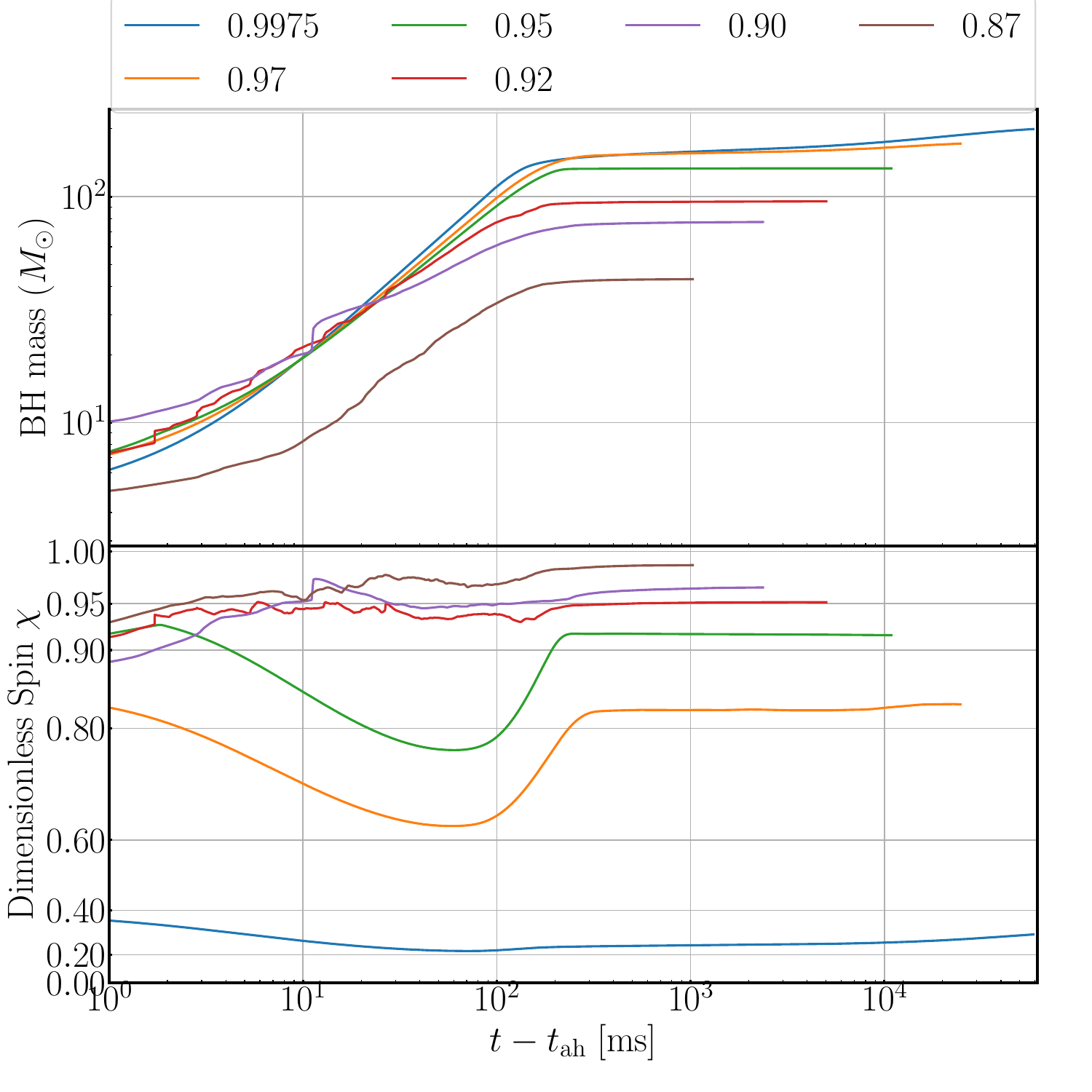}
    \includegraphics[width=0.49\textwidth]{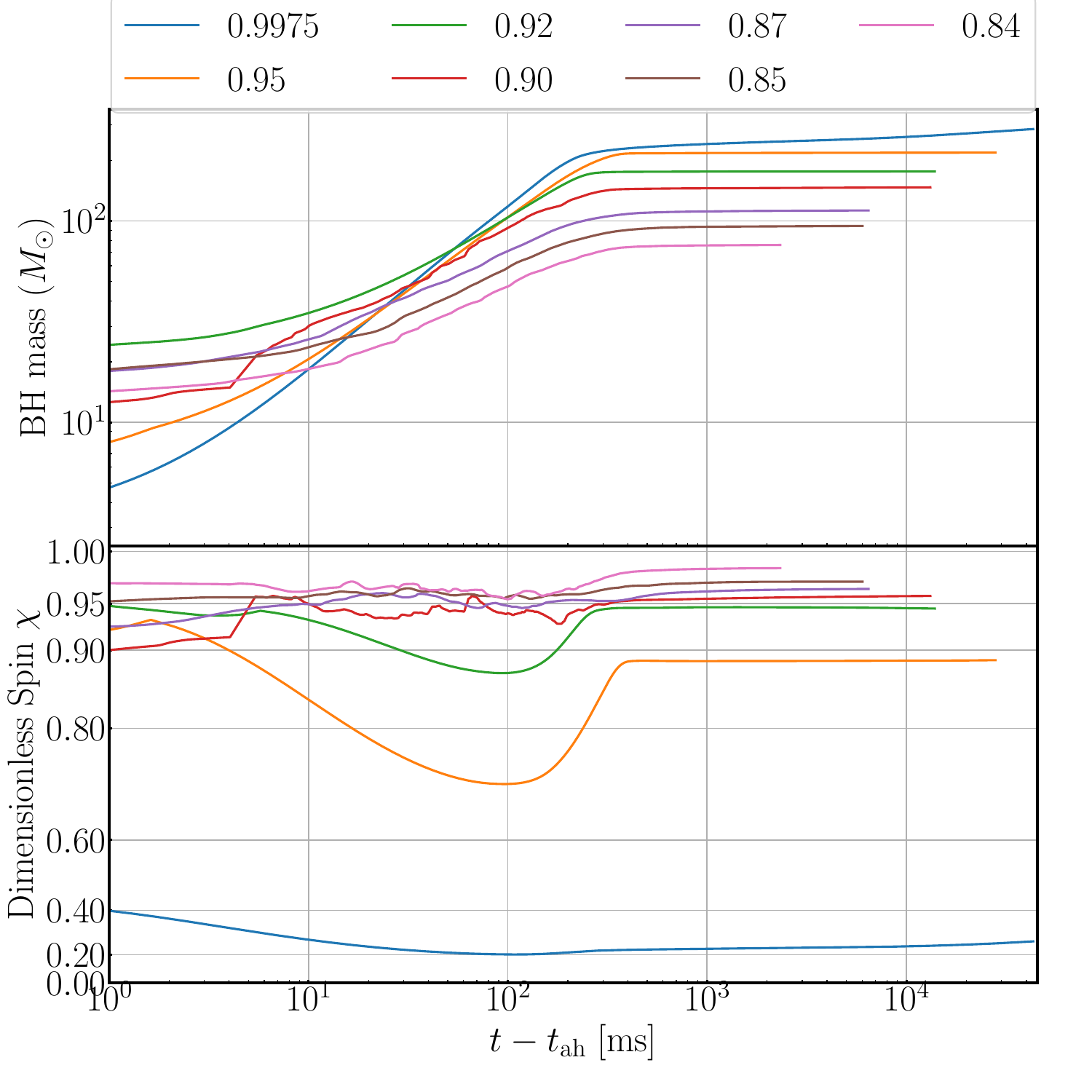}\\
    \includegraphics[width=0.49\textwidth]{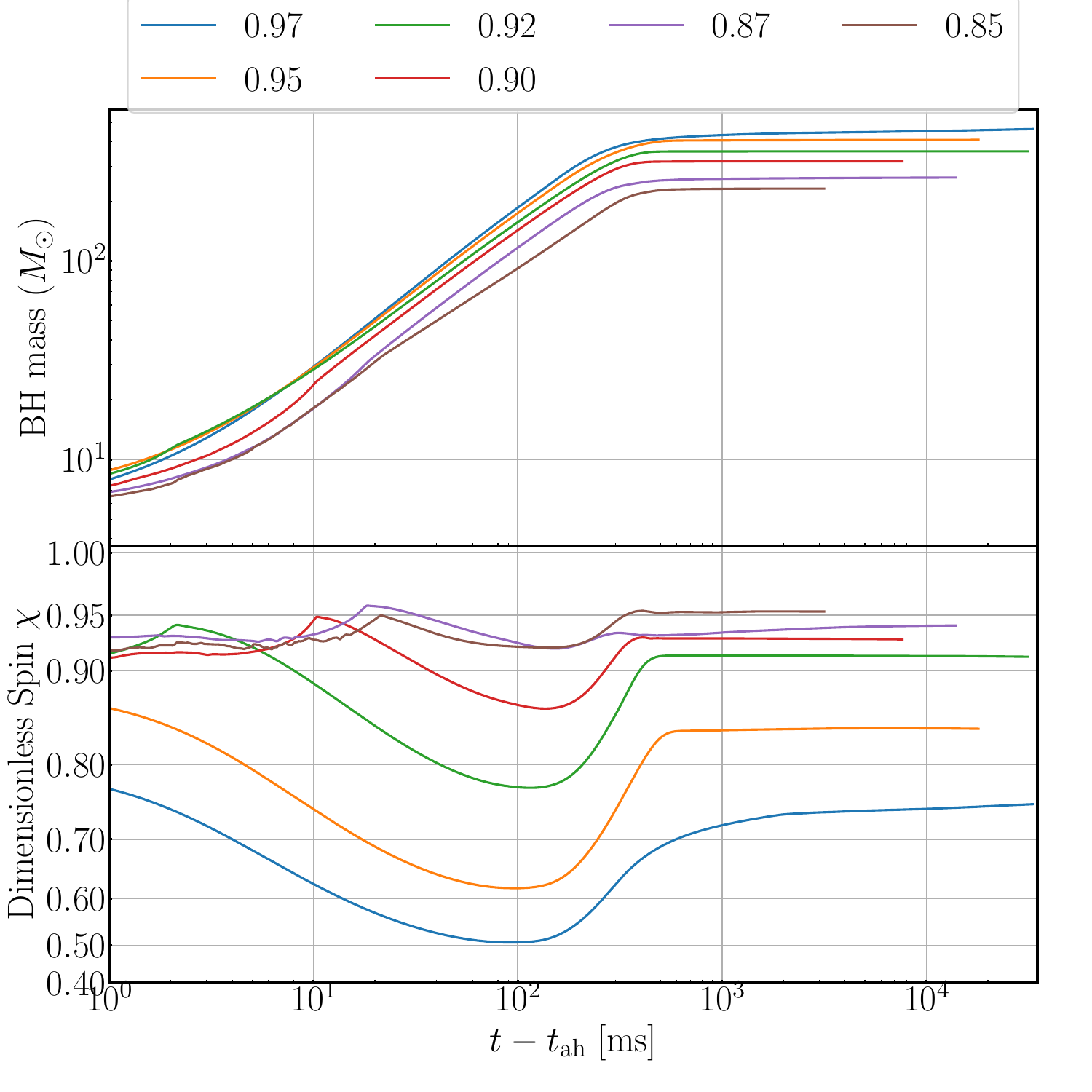}
    \includegraphics[width=0.49\textwidth]{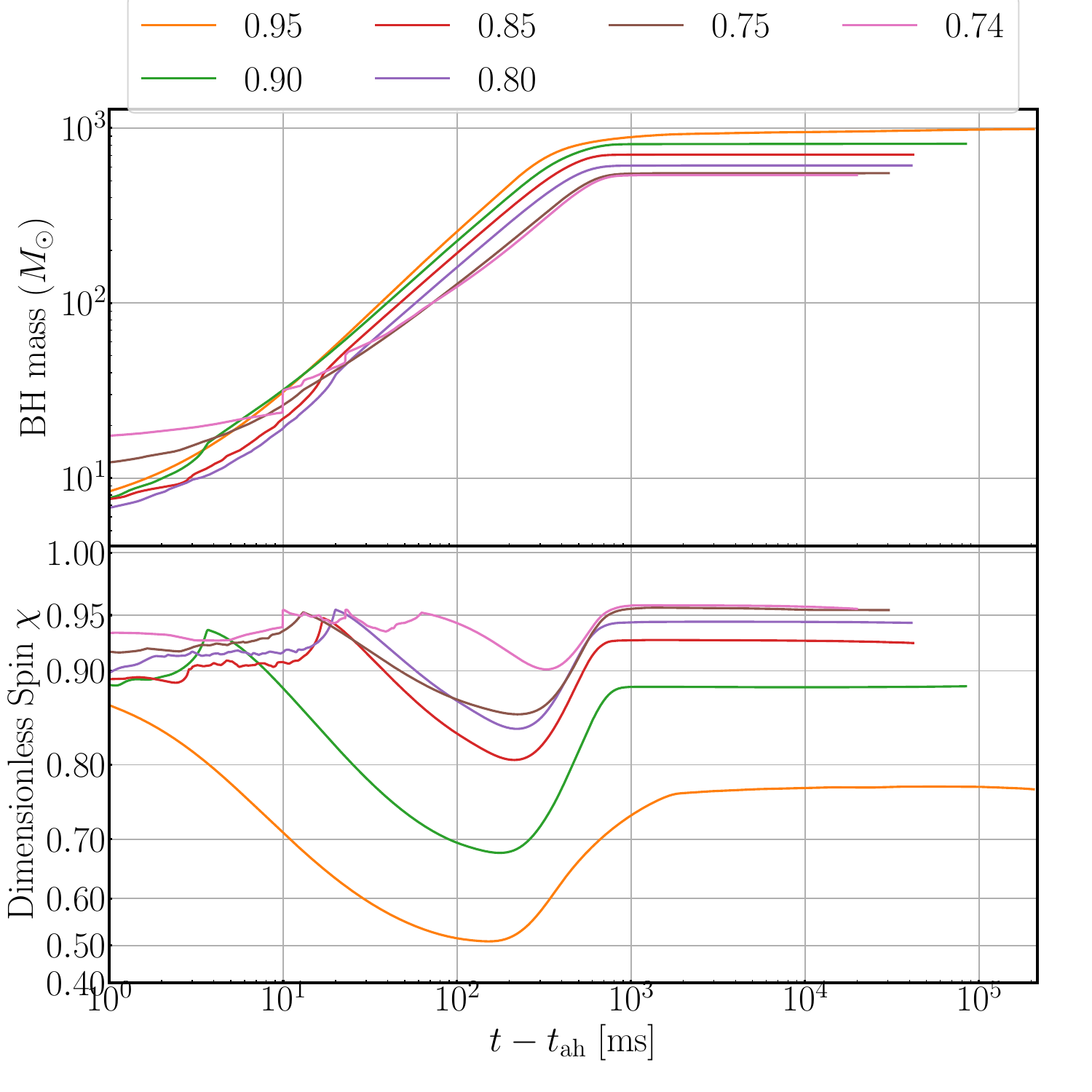}
    \caption{Evolution of the black hole mass and dimensionless spin after apparent horizon formation for models S17 (top left), S20 (top right), S25 (bottom left), and S35 (bottom right). Note that the total mass of the system is $\approx 200$, 300, 500, and $1100M_\odot$ for models S17, S20, S25, and S35, respectively. $t_\mathrm{AH}$ denotes the time at the first formation of the apparent horizon. 
    }
    \label{fig_bh_evo}
\end{figure*}

\begin{figure}
    \centering
    \includegraphics[width=0.95\columnwidth]{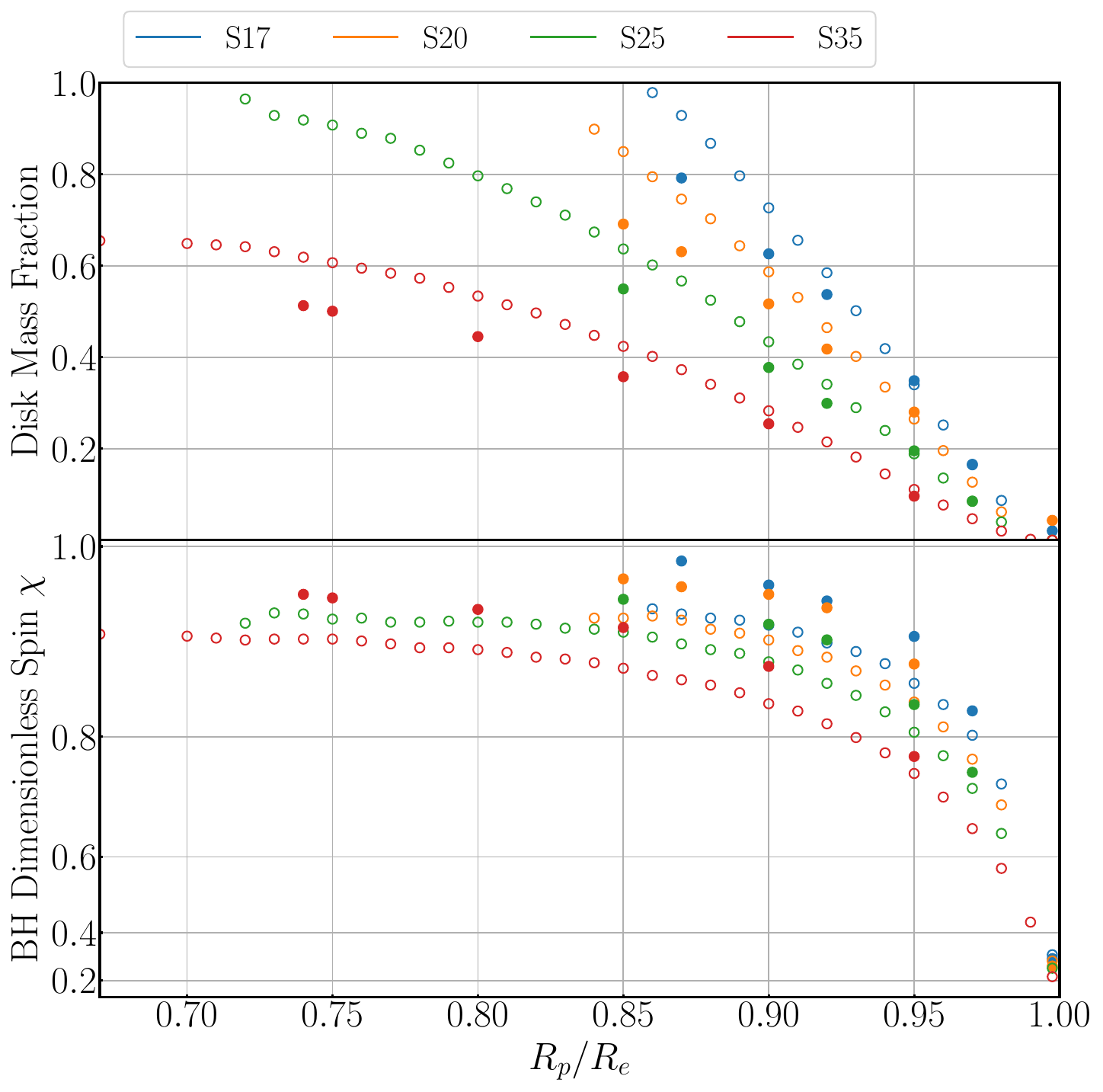}
    \caption{Filled markers: Disk mass fraction and dimensionless spin of black holes measured at the end of each simulation for models S17 (green), S20 (orange), S25 (green), and S35 (red). Hollow markers: Prediction from the initial data made in Ref.~\cite{Shibata:2025lde}. 
    }
    \label{fig_bhdisk}
\end{figure}

The evolution process of class (i) collapse can be further divided into two categories. For models with slower rotation, the collapse continues until a black hole forms without halting. Conversely, if the rotation is relatively fast, a centrifugal bounce happens before a black hole forms, at a maximum density of less than $10^{14}\,\mathrm{g/cm^3}$.
Figure~\ref{fig_rhomax_ah} illustrates how the maximum density, $\rho_\mathrm{max}$, evolves over the 100\,ms preceding black hole formation. Models with $R_\mathrm{p}/R_\mathrm{e}$ near unity lead to direct black hole formation without a significant bounce. On the other hand, models with smaller $R_\mathrm{p}/R_\mathrm{e}$ undergo a bounce when the maximum density reaches $10^{13}$--$10^{14}\,\mathrm{g/cm^3}$, a value lower than the nuclear saturation density, as previously noted. 
After this bounce, a dense, flat spheroid is temporarily formed, with a lifetime of several 10\,ms. 

Figure~\ref{fig_2dcont} displays snapshots of the density profiles for model S20 with $R_\mathrm{p}/R_\mathrm{e}=0.85$ at selected time slices around the black hole formation time. For this model, a dense, flat spheroid with a lifetime of $\sim 80$\,ms forms (see the first--third panels). The central region of the dense spheroid subsequently collapses into a black hole after accreting mass from the outer region, and the black hole then gradually grows with time (see the fourth--sixth panels). Matter surrounding the central region is accreted by the black hole as it grows, but an extended disk forms (see the seventh--ninth panels at a larger scale). This disk is flat, massive, and therefore susceptible to non-axisymmetric instabilities (see Sec.~\ref{sec4.4}). We note that the dense, flat spheroids did not exhibit high-amplitude density oscillations, and thus the evolution process differs from that of class (ii). 

In most models, the bounce results from centrifugal force and a slight stiffening of the equation of state due to neutrino trapping, rather than the stiffening of the equation of state in nuclear matter. This justifies our simplified approach to the nuclear equation of state (see also Refs.~\cite{2007ApJ...666.1140N, Sekiguchi2011} for related discussions). Nonetheless, it is important to note that the evolution of the dense spheroid may be influenced by the equation of state, warranting more detailed investigation to accurately determine its behavior. This aspect will be addressed in future research. 

We observe non-systematic behavior in model S17 with $R_\mathrm{p}/R_\mathrm{e}= 0.87$ and model S20 with $R_\mathrm{p}/R_\mathrm{e}=0.84$, both lied at the boundary between classes (i) and (ii). For these models, the dense spheroid that forms temporarily is shorter than in models with slightly smaller $R_\mathrm{p}/R_\mathrm{e}$, although even at larger $R_\mathrm{p}/R_\mathrm{e}$, the spheroid lifetime anti-correlates with $R_\mathrm{p}/R_\mathrm{e}$. This pattern occurs because, in these special cases, a weak bounce causes a short-term deceleration of the central density increase when $\rho_\mathrm{max}$ nears $\rho_1$ (see insets of Fig.~\ref{fig_rhomax_2d}). Consequently, the collapse experiences a slight delay at $\rho=\rho_\mathrm{max} \sim \rho_1$, and during this delay, mass accumulates toward the center. After this weak bounce, the collapse accelerates with more mass concentrated in the central region, shortening the dense spheroid's lifetime compared to models with slightly lower angular momentum. 

\subsection{Mass and spin of black holes and disk mass}\label{sec4.3}

Figure~\ref{fig_bh_evo} shows the evolution of the mass and dimensionless spin of the black holes (denoted by $M_\mathrm{BH}$ and $\chi_\mathrm{BH}$) for selected models explored in axisymmetric simulations. Here, the black hole quantities are determined from the area and circumferential radii of the apparent horizons (see, e.g., Ref.~\cite{Shibata2016a}). For all the models, the black hole mass at formation, $M_\mathrm{BH,0}$, is significantly smaller than the total mass of the system, typically around $M_\mathrm{BH,0} \sim 5$--$20M_\odot$, because the collapse proceeds in a runaway manner, and thus only the central, high-density region collapses to form a black hole in the early stage. As emphasized in Ref.~\cite{Fujibayashi:2026}, this is different from that of the collapse of supermassive stars with mass $\agt 10^4M_\odot$ (see, e.g., Refs.~\cite{Shibata:2002br, 2007PhRvD..76h4017L, Uchida:2017qwn, Fujibayashi:2024vnb}), for which the collapse is triggered by the f-mode radial-oscillation instability and proceeds in a rather homologous manner. 

Subsequently, the black hole mass increases steeply and stabilizes at about $\sim 10^3GM/c^3$. For models with slow rotation, the black hole accretes a substantial portion of the total mass, $M$. In contrast, for rapidly rotating cases, the relaxed black hole mass is less than half of the total mass, as predicted in Ref.~\cite{Shibata:2025lde}. In these cases, an appreciable fraction of the matter remains outside the black hole, leading to the formation of a massive disk, and the dimensionless spin is high, with $\chi_\mathrm{BH} \agt 0.95$; a rapidly spinning black hole forms (see the filled symbols in Fig.~\ref{fig_bhdisk}). In particular, 
for the case that $M_\mathrm{BH}/M \alt 2/3$ (i.e., for $R_\mathrm{p}/R_\mathrm{e} \alt 0.95$, 0.92, and 0.85 for models S17, S20, and S35), a massive disk can develop. We conduct 3D simulations for these scenarios. 

We note that the results presented in Figs.~\ref{fig_bh_evo} and \ref{fig_bhdisk} are based on axisymmetric simulations. In 3D simulations, where non-axisymmetric instabilities develop in the massive disk, the further infall of matter onto the black hole proceeds via angular-momentum transport associated with the instability. Thus, the final mass and dimensionless spin of the black holes are slightly modified (see Sec.~\ref{sec4.3}). 

\begin{table*}[t]
\caption{List of the selected models for 3D simulations. The third--ninth columns show $R_\mathrm{p}/R_\mathrm{e}$, initial core mass, black hole mass and dimensionless spin at the initial formation of apparent horizon (denoted by $M_\mathrm{BH,0}$ and $\chi_\mathrm{BH,0}$), $M_\mathrm{BH}$, $\chi_\mathrm{BH}$, and the approximate peak frequency of gravitational waves of the $l=m=2$ mode. 
$t_{\rm 3D, sta}$ indicates the starting time of the 3D simulations after the AH formation.
The eleventh to fifteenth columns show the signal-to-noise ratio for gravitational waves at a hypothetical distance of 3\,Gpc with respect to the designed sensitivity curves of the advanced LIGO, the Einstein Telescope (ET-B, ET-D), and the Cosmic Explorer (CE1, CE2). $M_\mathrm{BH}$ and $\chi_\mathrm{BH}$ are evaluated at 1\,s after the formation of the black holes. For all the listed models, massive disks formed around the black hole become unstable. 
\label{tab1}
}
\begin{tabular}{ccccccccccccccc}
\hline
Model & $s/k$  & $R_\mathrm{p}/R_\mathrm{e}$ & $M/M_\odot$& $M_\mathrm{BH,0}/M_\odot$ & $\chi_\mathrm{BH,0}$
& $M_\mathrm{BH}/M_\odot$ & ~$\chi_\mathrm{BH}$~ &  $f_{22}$\,(Hz)
& $t_{\rm 3D, sta} ({\rm ms})$
& aLIGO & ET-B & ET-D & ~CE1 & ~CE2 \\ \hline
S17  & 17  & $0.90$ &206.9& 8.09 & 0.848 & 77.3 & 0.966 & 39 &1.99 &0.3&3.9&4.3&7.7&14\\ \hline
S17  & 17 & $0.95$ &205.2& 4.62 & 0.847 & 134 & 0.917 & 47 & 6.70 &0.2&2.3&2.2&4.4&7.9\\\hline   
S20  &20  & $0.85$ &306.3& 14.3 & 0.922 & 94.6 & 0.972 & 36 &15.74 &0.4 & 5.6&9.3&13&25\\ \hline
S20 & 20  & $0.92$ &303.3& 21.2  &  0.950  & 177 & 0.945 &  34& 79.99 & 0.2&3.2&4.8&8.6&16\\\hline
S25  & 25 & $0.85$ &514.0& 5.04 & 0.890 & 232  & 0.953 & 34 & 2.93 &0.9 & 13 & 15 & 29 & 51 \\\hline
S25  & 25 & $0.90$ &510.7& 5.46 & 0.879 & 318 & 0.929 & 21 & 7.53 &0.2&2.8&8.8&12&25\\ \hline
S35  & 35 & $0.75$ &1106& 10.6 & 0.924 & 552 & 0.954 & 13 & 1.65 &0.2&4.1&31&11&37\\ \hline  
S35  & 35 & $0.90$ &1090& 6.15 & 0.869 & 813 & 0.885 & 9.5 & 10.45 &0.07&1.2&6.7&3.8&7.9\\ \hline 
\end{tabular}
\end{table*}

Before presenting the 3D simulation results, we compare the black hole mass and dimensionless spin with the predicted values from the initial data, by examining the filled (simulation results) and hollow (predictions) markers in Fig.~\ref{fig_bhdisk}. The upper and lower panels show the fraction of the mass of the matter located outside the black hole (referred to as the disk mass fraction) and the dimensionless spin of the black hole $\chi_\mathrm{BH}$, respectively. This figure shows that the prediction can overestimate the disk mass fraction and underestimate the dimensionless spin, although the errors remain within 20\% for the disk mass fraction and $\sim 5\%$ for the dimensionless spin. Therefore, we can say that the prediction based on the initial data is quite accurate. A potential reason for this systematic error is that, in our prediction, we assumed that matter with lower specific angular momentum would collapse into black holes sequentially. Specifically, we considered that matter falls into the black hole if its specific angular momentum, $j$, is less than that of the innermost stable circular orbit, $j_\mathrm{ISCO}$, which depends on the black hole's mass and spin over time. However, the actual collapse dynamics are more complex; for example, matter falling in may follow highly elliptical orbits due to radial infall. In such cases, even when $j > j_\mathrm{ISCO}$, matter can still fall into the black hole, increasing its spin and reducing the mass of matter outside the black hole. 

\subsection{Results of 3D simulations}\label{sec4.4}

\begin{figure*}
    \centering
    \includegraphics[width=\textwidth]{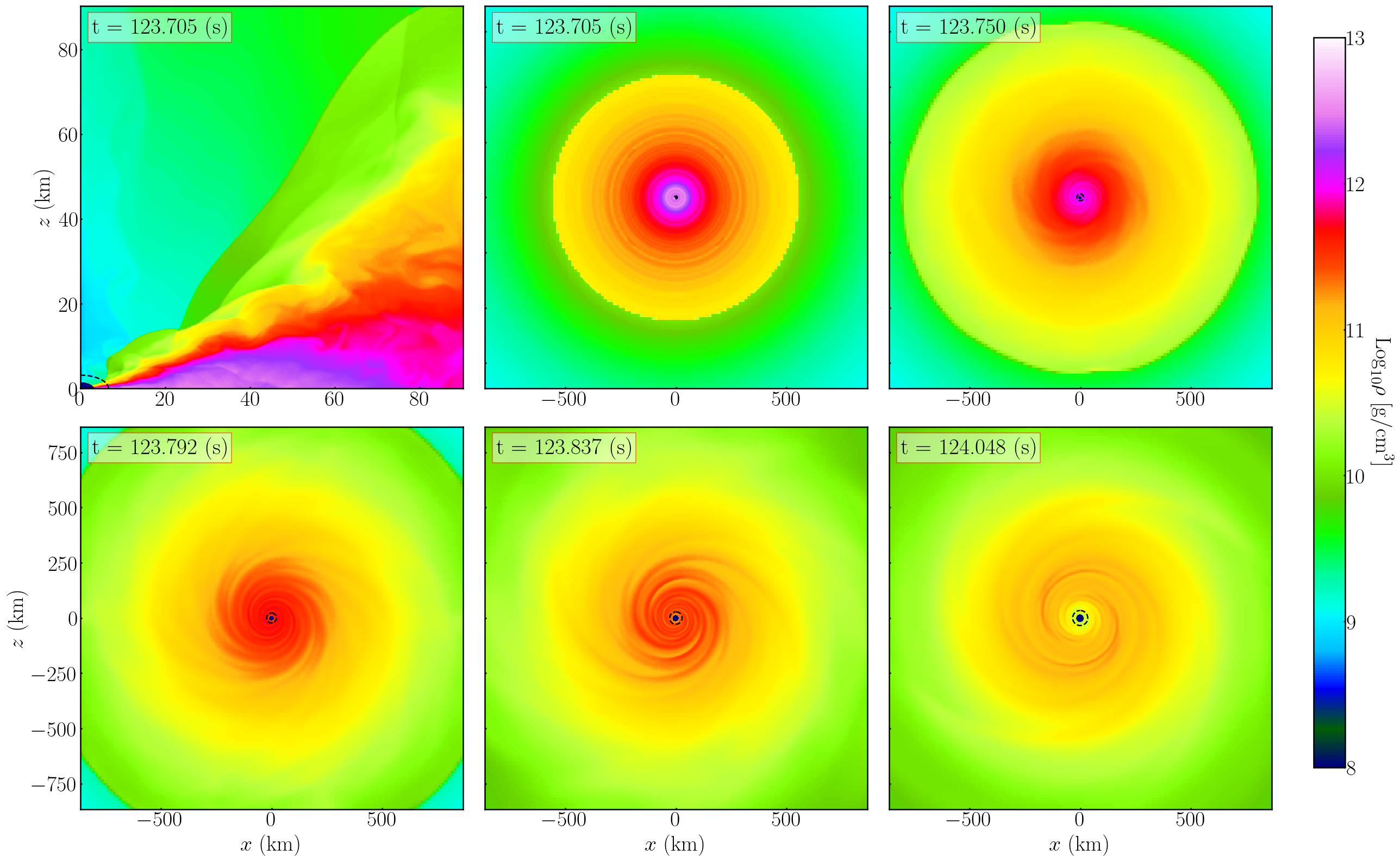}
    \caption{Snapshots of the density profiles in the 3D simulations at selected time slices for model S20 with $R_\mathrm{p}/R_\mathrm{e}=0.85$. The first panel is a snapshot near the black hole in the $x$-$z$ plane at the start of the 3D simulation ($\approx 15$\,ms after the black hole's formation), while the other 5 panels show the density profiles in the equatorial plane. The time in the second panel is the same as in the first panel. The filled and dashed circles around the origin denote the computationally excised region and apparent horizon, respectively. The density is shown in units of $\mathrm{g/cm^3}$. An animation for the 3D evolution is found at \url{https://www2.yukawa.kyoto-u.ac.jp/~masaru.shibata/movierho_xy_20.085.mp4}.
    }
    \label{fig_3dcont}
\end{figure*}

\begin{figure}
    \centering
    \includegraphics[width=0.95\columnwidth]{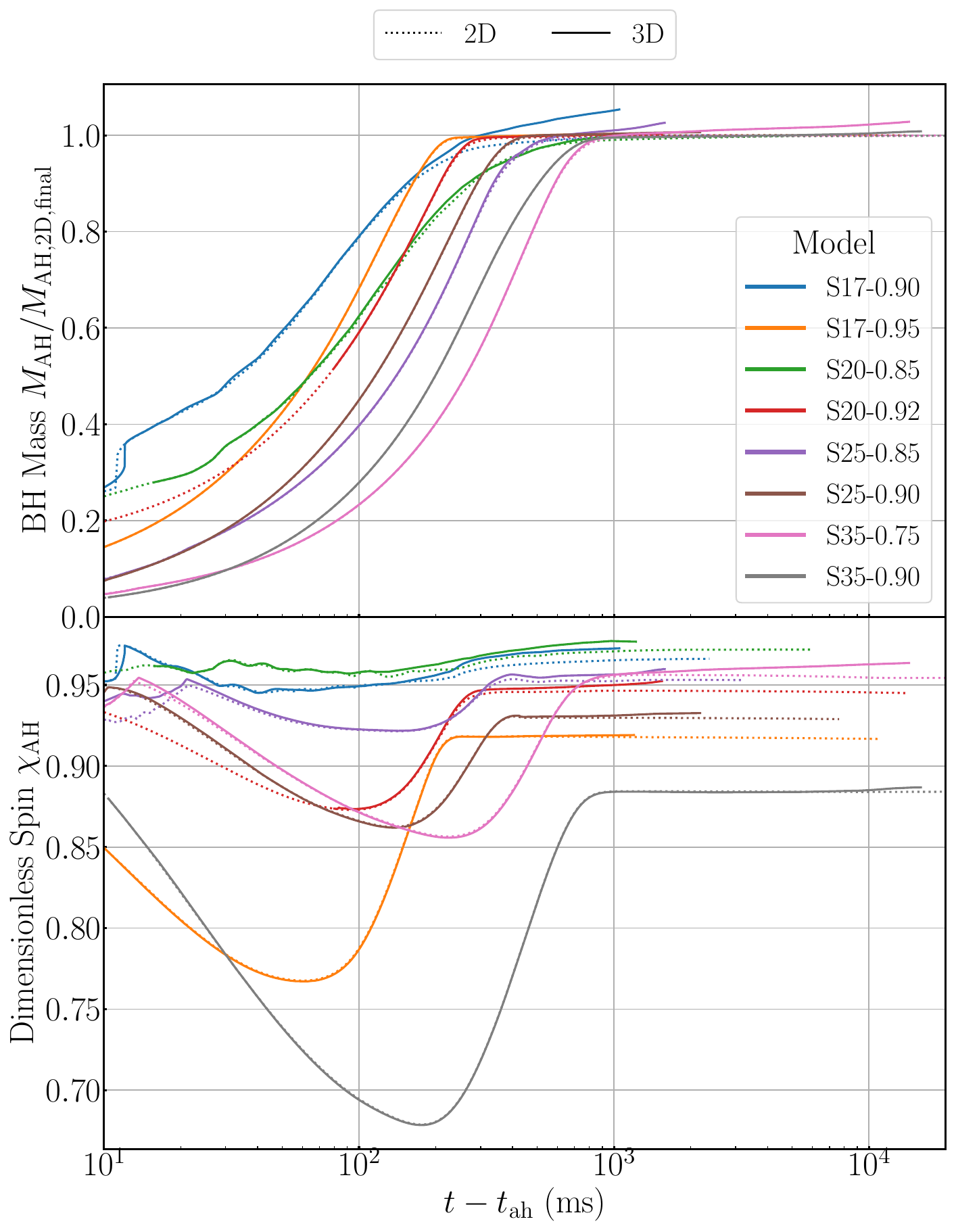}
    \caption{The evolution for the black hole mass (upper panel) and dimensionless spin (lower panel) for the axisymmetric (dotted) and the corresponding 3D (solid) models. In the upper panel, the black hole mass is normalized by the final black hole mass in the axisymmetric simulations. 
    }
    \label{fig_ah_mass_spin_2d3d}
\end{figure}

\begin{figure}
    \centering
    \includegraphics[width=0.95\columnwidth]{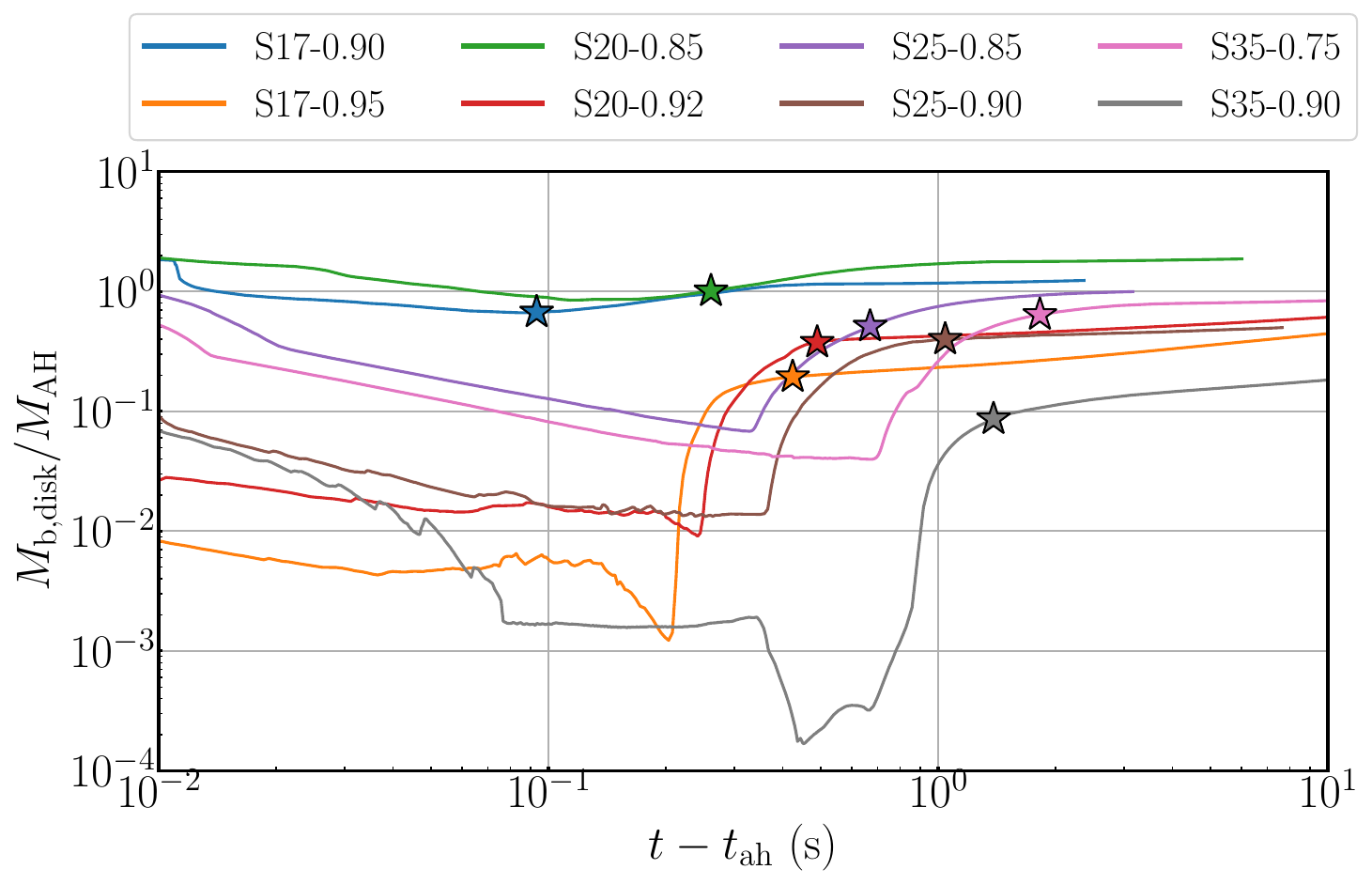}
    \caption{The ratio between the disk baryon mass and the apparent horizon mass as a function of time. The star markers show the moment when the gravitational wave amplitude peaks.}
    \label{fig_disk_mass_ratio}
\end{figure}

We conduct 3D simulations for specific models (see Table~\ref{tab1}), focusing especially on cases with moderate rotation speeds, where $R_\mathrm{p}/R_\mathrm{e}=0.85$--0.95. For all these models, a rapidly spinning black hole with $\chi_\mathrm{BH} > 0.85$ forms alongside a massive disk. As mentioned in Sec.~\ref{sec2}, the initial data for the 3D simulations are taken from axisymmetric simulations carried out in advance. The data are imported during the black hole’s early growth phase, when its mass is still much smaller than the core mass, $M$. Due to computational limitations, we do not attempt to resolve black holes immediately after their formation, as their masses are tiny and resolving them is computationally expensive. The 3D simulations are initiated once the black hole reaches $\gtrsim 10M_\odot$, typically 0--80 ms after black hole formation (see $t_\mathrm{3D,sta}$ in Table~\ref{tab1}). Therefore, in this study, we do not analyze the non-axisymmetric stability of the disks immediately following black hole formation.

In all cases presented in Table~\ref{tab1}, the massive disks surrounding the black holes are susceptible to non-axisymmetric deformation. The instability is developed until the deformation becomes non-linear. Nonetheless, the extent of this instability varies with the models' mass and rotational properties, as these factors influence the disk mass (or the disk-to-black-hole mass ratio).

Figure \ref{fig_3dcont} illustrates density contour plots at chosen time slices for model S20 with $R_\mathrm{p}/R_\mathrm{e}=0.85$. This model develops a long-lived dense spheroid before a black hole forms (see Fig.~\ref{fig_rhomax_ah}). As a result, a dense disk with a maximum density exceeding $10^{12}\,\mathrm{g/cm^3}$ has already existed at the moment of black hole formation. In contrast, models with higher values of $R_\mathrm{p}/R_\mathrm{e}$ (i.e., lower stellar core angular momentum) show different behavior. For example, for model S20 with $R_\mathrm{p}/R_\mathrm{e}\geq 0.90$, the dense spheroid is short-lived, and the maximum density and disk mass at black hole formation are lower than in the 0.85 model. Higher $R_\mathrm{p}/R_\mathrm{e}$ models thus experience a delay in the onset of non-axisymmetric instability until the disk gains mass from continued infall. Nevertheless, the growth timescale remains around 0.1 seconds, comparable to the orbital periods near the innermost stable circular orbits of massive black holes, indicating a dynamical instability. This is also evident in the gravitational waveforms (see Fig.~\ref{fig_gw} in the next section). We note that the dynamical timescale is much shorter than other timescales such as the viscous timescale, making it unlikely that dissipative mechanisms will significantly inhibit the non-axisymmetric deformation (see Refs.~\cite{2018MNRAS.475..108B,Ruiz:2026ytb}).

In all cases, the disk is very compact, the density peak is located at $\sim 1.5GM_\mathrm{BH}/c^2$ in quasi-isotropic coordinates (see the first panel in Fig.~\ref{fig_3dcont}). This compactness is due to the black holes' rapid spin, which allows such a compact orbit. When the disk becomes non-linearly unstable to a non-axisymmetric instability, multiple spiral arms develop (see the third--fifth panels).
This instability transports angular momentum within the disk through an enhanced gravitational torque, thereby reactivating mass accretion onto the black hole, leading to higher black hole mass and dimensional spin than in axisymmetric simulations (see Fig.~\ref{fig_ah_mass_spin_2d3d}). As a result of angular momentum redistribution, the disk expands, and the maximum density and disk mass (for the high-density part) decrease with time. The disk subsequently becomes less unstable, although a weak non-axisymmetric perturbation persists for a long period.

In higher-mass models like the S35, the instability is less pronounced at a given $R_\mathrm{p}/R_\mathrm{e}$ ratio (e.g., 0.90) because of the lower disk-to-black-hole mass ratio. For the S35 model with $R_\mathrm{p}/R_\mathrm{e}=0.90$, this ratio remains around 0.2 even at the final stage, and the disk's maximum density is typically about $10^8$ g/cm³, significantly lower than in the S20 and S25 models. Although the non-axisymmetric instability, primarily the bar-mode, is present, its perturbation amplitude is smaller than in the S20 and S25 models. Additionally, the non-axisymmetric instabilities develop later in the S35 model. This suggests that other efficient angular momentum transport processes, such as the magnetorotational instability~\cite{Balbus:1998ja, 2018MNRAS.475..108B}, could further weaken the disk's non-axisymmetric instability. 

However, for smaller $R_\mathrm{p}/R_\mathrm{e}$ values, which produce more rapidly spinning black holes, the disk tends to be more massive, more compact, and higher density. Consequently, the pattern of non-axisymmetric deformation resembles that seen in models S20 and S25 with $R_\mathrm{p}/R_\mathrm{e}=0.85$. For example, in model S35 with $R_\mathrm{p}/R_\mathrm{e}=0.75$, the disk mass is higher (see Fig.~\ref{fig_disk_mass_ratio}), resulting in a much greater degree of non-axisymmetric instability than in the S35 model with $R_\mathrm{p}/R_\mathrm{e}=0.90$. These differences in instability dynamics are also reflected in the gravitational wave amplitudes. 

As observed in our previous work~\cite{Kiuchi:2011re}, the disk becomes unstable when the disk-to-black-hole mass ratio is sufficiently high. To separate the disk from the surrounding infalling matter, we define the disk as the region where the specific thermal energy $\varepsilon - \varepsilon_{\rm cold}$ exceeds 10\% of the specific internal energy $\varepsilon$. Figure~\ref{fig_disk_mass_ratio} illustrates how this mass ratio evolves over time across all 3D models. The star markers show the moment when the gravitational wave amplitude peaks. This figure shows that for models with an initial mass ratio above $\sim 1$ at the start of the 3D simulation (model S17 with $R_\mathrm{p}/R_\mathrm{e}=0.90$ and model S28 with $R_\mathrm{p}/R_\mathrm{e}=0.85$), the non-axisymmetric instability occurs within 0.1 seconds of the start of the 3D simulation (cf. the 3D gravitational waveforms in the following subsection). In contrast, when the initial mass ratio is below 0.1, the development of non-axisymmetric instability is delayed. Examples include the S20 model with $R_\mathrm{p}/R_\mathrm{e}=0.92$, the S25 model with $R_\mathrm{p}/R_\mathrm{e}=0.90$, and the S35 model with $R_\mathrm{p}/R_\mathrm{e}=0.90$. In these cases, the instability develops only after the mass ratio surpasses $\sim 0.1$. The disk-to-black-hole mass ratio eventually approaches an asymptotic value. The peak amplitude of gravitational waves is typically recorded when the ratio starts approaching the asymptotic values.

\subsection{Gravitational waves}

The collapse of rotating, very-massive stars can produce burst sources of gravitational waves, particularly when a massive disk forms and becomes unstable to non-axisymmetric deformation. During disk and black hole formation, burst gravitational waves, along with disk oscillations and black hole ringdown, may be emitted. This subsection discusses these gravitational waves. Before presenting numerical results, we summarize the characteristic frequencies and amplitudes expected for various gravitational wave modes. We focus only on gravitational waves with frequencies between 1 Hz and a few kHz, as these are relevant to ground-based detectors such as Advanced LIGO, ET, and CE.

Assuming that a non-axisymmetrically deformed disk has the maximum density around a radius close to the black hole, the characteristic frequency of gravitational waves is estimated as
\begin{align}
f&={\Omega \over \pi}
\approx 41\,\mathrm{Hz} \left(\frac{\varpi}{3GM_\mathrm{BH}/c^2}\right)^{-3/2}
\left(\frac{M_\mathrm{BH}}{300M_\odot}\right)^{-1}, \label{eq13}
\end{align}
where $\varpi$ denotes the typical cylindrical radius of the disk, assuming the disk has circular orbits. Therefore, for $M_\mathrm{BH}\alt 10^3M_\odot$, the frequency is likely to exceed $\sim 10$\,Hz, and the emitted gravitational waves can be a source for ground-based detectors if the disk is sufficiently compact.

When a massive disk forms around a black hole, the black hole tends to spin rapidly, as shown in earlier sections. This results in a more compact disk because the radius of the innermost stable circular orbit (in units of $GM_\mathrm{BH}/c^2$) decreases with increasing spin~\cite{Bardeen:1972fi, 1983bhwd.book.....S}. Consequently, the gravitational wave frequency can be higher. This property is also suitable for detecting gravitational waves from very-massive stellar core collapses that form massive black holes, since ground-based detectors are most sensitive to frequencies above about 10 Hz~\cite{Hild:2010id,Punturo2010}. 

The amplitude of gravitational waves is approximately given by the quadrupole formula as
\begin{equation}
h \sim \left|{G(\ddot I_{yy}-\ddot I_{xx}) \over c^4 D}\right|,
\end{equation}
where $D$ denotes the (luminosity) distance to the source, $I_{ij}$ is the quadrupole moment, and $\ddot I_{ij}=d^2I_{ij}/dt^2$. Estimating $I_{yy}-I_{xx}$ and $\ddot I_{yy}-\ddot I_{xx}$ by
\begin{align}
\begin{split}
|I_{yy}-I_{xx}| &\sim \epsilon M_\mathrm{disk} \varpi^2,\\
|\ddot I_{yy}-\ddot I_{xx}| &\sim 4 \epsilon M_\mathrm{disk} \varpi^2 \Omega^2,    
\end{split}
\end{align}
where $\Omega$ is the typical orbital angular velocity, $\varpi$ is the typical cylindrical radius of the disk, and $\epsilon$ denotes the fraction of the disk mass that contributes to the non-axisymmetric oscillation, we obtain
\begin{align}
h \sim {4G \epsilon M_\mathrm{disk} \varpi^2 \Omega^2 \over c^4D} 
 &\approx 5 \times 10^{-23}\,\left({\epsilon \over 0.1} \right)
 \left({M_\mathrm{disk} \over 10 M_\odot}\right) \nonumber \\
 \times&
\left({GM_\mathrm{BH}/c^2\varpi \over 0.25}\right)
\left({D \over 1\,\mathrm{Gpc}}\right)^{-1},
\label{eq19}
\end{align}
where we assumed $\Omega^2 \approx GM_\mathrm{BH}/\varpi^3$. 

After a spinning black hole forms, ringdown gravitational waves can be emitted. The perturbative analysis of spinning black holes~\cite{Berti:2009kk} gives the frequency of the axisymmetric ringdown mode as
\begin{eqnarray}
f_\mathrm{RD} &\approx& 3.23\,\mathrm{kHz}  
\left({M_\mathrm{BH} \over 10M_\odot}\right)^{-1} \nonumber \\
&&\times \left[0.4437-0.0739(1-\chi_\mathrm{BH})^{0.3350}\right]. \label{eq20}
\end{eqnarray}
This implies that for $\chi_\mathrm{BH}=0.9$ and $0.95$, $f \approx 1.32$ and $1.35$\,kHz, respectively, for $M_\mathrm{BH}=10M_\odot$. Here, we assume a relatively small black hole mass because the ringdown gravitational waves are emitted in the early stage of black hole growth. The frequency also falls within the optimal band for ground-based detectors for black holes of orders $10$--$10^3M_\odot$. It’s important to note that in the early evolution stages, the black hole mass is much less than $100 M_\odot$ (see Fig.~\ref{fig_bh_evo}), so the ringdown frequency can approach 1\,kHz, as we demonstrate below. The typical amplitude of gravitational waves for rapidly rotating progenitor cases is \cite{Shibata:2016vzw, Uchida:2019gjx} (see also the results below)
\begin{eqnarray} 
h &\sim& 4 \times 10^{-3} {GM_\mathrm{BH} \over c^2D} \nonumber \\
&\approx& 2 \times 10^{-23} \left({M_\mathrm{BH} \over 10^2M_\odot}\right)
\left({D \over 1\,\mathrm{Gpc}}\right)^{-1},
\end{eqnarray}
suggesting that ringdown gravitational waves could serve as sources for the ET and the CE, especially if emitted from high-mass systems with $M_\mathrm{BH} \gg 10M_\odot$, assuming $D \agt 1$ Gpc.

Furthermore, massive, dense disks can form around the black hole due to free fall in this scenario. As these disks form and grow, they undergo dynamic changes; consequently, high-amplitude gravitational waves might be emitted from their oscillations, even without non-axisymmetric deformation. If the frequency is roughly determined by the disk's dynamical timescale, the gravitational-wave frequency aligns with that in Eq.~\eqref{eq13}. Nonetheless, during the early stages of black hole growth, its mass can be significantly less than $200M_\odot$, leading to higher frequencies. 
The amplitude of gravitational waves for the $l=2$, $m=0$ mode is approximately estimated by the quadrupole formula as
\begin{equation}
h \sim \left|{G(\ddot I_{zz}-\ddot I_{xx}) \over c^4 D}\right|.
\end{equation}
In this case, gravitational waves are induced by disk oscillations, and thus the characteristic oscillation timescale is $\sim H/c_\mathrm{s}$, where $H$ is the disk scale height and $c_\mathrm{s}$ is the sound speed. Assuming the force balance in the vertical direction of the disk, we have $H\Omega \sim c_\mathrm{s}$. 
Then, the order of magnitude of $h$ becomes comparable to Eq.~\eqref{eq19} if $\epsilon$ is of the same order of magnitude. 
We note that gravitational waves for the $l=2$, $m=0$ mode are emitted most strongly toward the equatorial plane, while those for the $l=m=2$ mode are emitted toward the rotational axis ($z$-axis), suggesting that gravitational waves with appreciable amplitude may be expected in any direction.

All these predictions are confirmed in the present numerical results. In what follows, we show waveforms for axisymmetric and 3D simulations separately. We pay attention to gravitational waves for models with moderately rapid rotation for which $R_\mathrm{p}/R_\mathrm{e} \leq 0.95$. Gravitational waves from less rapidly rotating models have a smaller amplitude. 

\subsubsection{Analysis methods}

As in standard numerical relativity simulations~(e.g., Ref.~\cite{Shibata2016a}), gravitational waves are extracted by computing the outgoing component of the complex Weyl scalar from the geometric quantities and decomposed into multipole components, $\Psi_{lm}$, using the spin-weighted spherical harmonics, $_{-2}Y_{lm}(\theta, \varphi)$. Fourier and inverse Fourier transforms are performed with a low-frequency filter~\cite{Reisswig:2010di}, as is standard in numerical relativity. We focus only on $(l, m)=(2,0)$ and $(2,\pm 2)$ for axisymmetric and 3D simulations, respectively. 

The Fourier spectrum is defined by 
\begin{align}
h_{lm}(f)&=\int dt\, h_{lm}(t) e^{-2\pi i f t} \nonumber \\
&\approx-\int dt\, {\Psi_{lm}(t) \over (2\pi)^2(f^2+f_\mathrm{cut}^2)} e^{-2\pi i f t}, 
\end{align}
where $f$ is the gravitational wave frequency, $f_\mathrm{cut}$ is a cutoff frequency for a low-frequency filter, and $h_{lm}(t)$ denotes the gravitational wave for each $(l,m)$ mode. $f_\mathrm{cut}$ is chosen to be much lower than the frequency at the Fourier peak. Unless otherwise stated, we will plot waveforms for $\theta=\pi/2$ and $0$ (i.e., the highest-amplitude direction) for axisymmetric and 3D simulation results, respectively. We define $h(f)=|h_{20}(f)_{-2} Y_{20}(\pi/2,\varphi)|$ for the axisymmetric case and $h(f)=|h_{22}(f)_{-2}Y_{22}(0, \varphi)+h_{2\,-2}(f)_{-2}Y_{2\,-2}(0, \varphi)|$ for the 3D case. 

For 3D simulations, the wave strain for each mode is calculated by
\begin{equation}
h_{lm}(t)=\int df \,h_{lm}(f) e^{2\pi i ft}.
\end{equation}
In the axisymmetric case, the waveform of $\Psi_{lm}(t)$ exhibits a memory-type deviation from zero asymptotically, presumably due to mass outflow. This makes obtaining a clear Fourier spectrum and strain using the previous method impractical; instead, we must employ an alternative approach, such as directly extracting the gravitational-wave strain via gauge-invariant wave extraction~\cite{Moncrief:1974am, Nagar:2006eu}. Therefore, for axisymmetric simulations, we only extract burst-like waves emitted near black hole formation (within $\sim \pm 3$--$5$\,ms with respect to formation time of the black hole) from the complex Weyl scalar, adjust the offset from zero artificially, and compute the Fourier spectrum without focusing on the long-term wave strain. We estimate the strain simply by focusing on the peak amplitude, as detailed below. 

Broadly speaking, the peak effective amplitude of gravitational waves may be estimated from $\Psi_{lm}$ using the following procedure. First, we extract the characteristic gravitational-wave frequency, $f_\mathrm{chr}$, from the wave pattern of $\Psi_{lm}(t)$, and then estimate the wave amplitude by
\begin{equation}
|h_{lm}(t)| \sim {|\Psi_{lm}(t)| \over (2\pi f_\mathrm{chr})^2}.
\end{equation}
Using the dimensionless quantity $\Psi_{lm}r_\mathrm{ex} M$, where $M$ is the total core mass and $r_\mathrm{ex}$ denotes the radius at which the complex Weyl scalar is extracted, we obtain a relation
\begin{equation}
h (D/M) \approx h_{lm} (r_\mathrm{ex}/M) \sim 
{\Psi_{lm} r_\mathrm{ex} M \over (2 \pi f_\mathrm{chr}M)^2}.
\end{equation}
Here, $h$ denotes the amplitude at an observer distance $D$, and $2 \pi f_\mathrm{chr} M\,(=2\pi Gc^{-3} f_\mathrm{chr} M)$ is dimensionless. For ringdown gravitational waves from rapidly rotating black holes, $f_\mathrm{chr}^{-1}$ is $\sim 15GM_\mathrm{BH}/c^3$ (cf. Eq.~\eqref{eq20}). Thus, the magnitude of $h D/M$ is approximately $(\Psi_{lm} r_\mathrm{ex} M)(2.4M_\mathrm{BH}/M)^2$ for ringdown gravitational waves. Since ringdown gravitational waves are emitted during the early evolution of black holes, $M_\mathrm{BH}/M$ is of order 0.1. Therefore, in the end, we will find that the gravitational-wave strain for the axisymmetric mode $h_{20}$ becomes smaller than that of $h_{22}$ associated with non-axisymmetric instabilities in the present context. 

We note that for large extraction radii, the high-frequency components are numerically dissipated due to poor resolution associated with the mesh refinement algorithm, and the amplitude can become spuriously smaller. Thus, the extraction radii are determined taking into account the wavelength of gravitational waves.
 
\subsubsection{Axisymmetric cases}

\begin{figure}
    \centering
    \includegraphics[width=0.98\columnwidth]{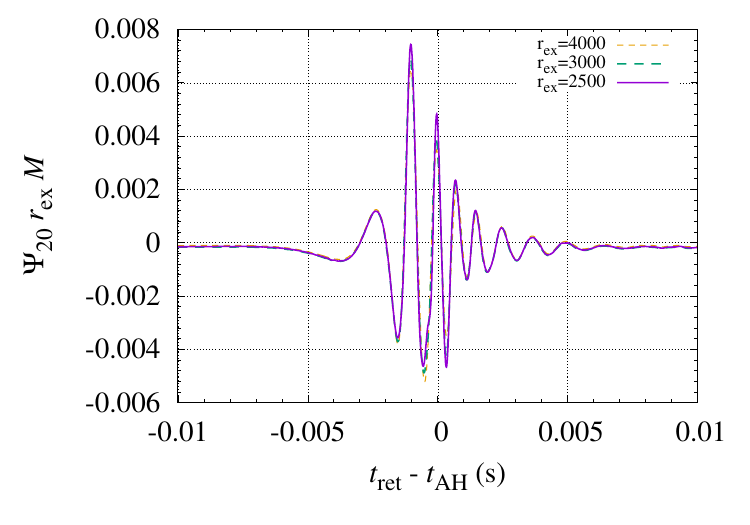}\\
    \vspace{-4mm}
    \includegraphics[width=0.98\columnwidth]{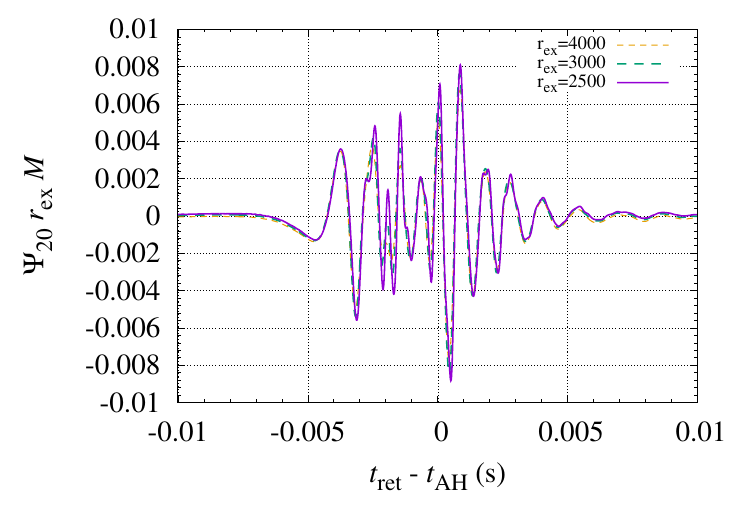}\\
    \vspace{-4mm}
    \includegraphics[width=0.98\columnwidth]{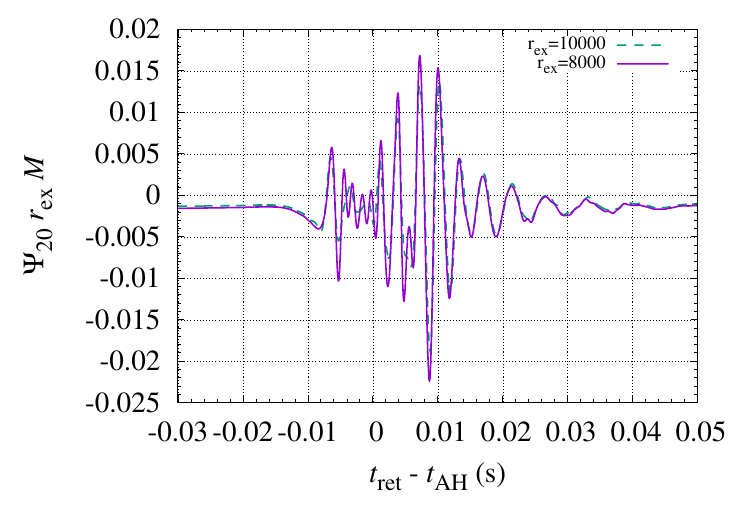}\\
    \vspace{-6mm}
    \caption{$\Psi_{20} r_\mathrm{ex}M$ as a function of a retarded time for models S17 (upper panel) and S20 (middle panel) with $R_\mathrm{p}/R_\mathrm{e}=0.95$, and model S35 with $R_\mathrm{p}/R_\mathrm{e}=0.80$ (bottom panel). We plot the raw numerical data of $\Psi_{20}$ with  different extraction radii. The extraction radii are shown in the upper right corner, in units of $GM_\odot/c^2 \approx 1.4767$\,km. The horizontal axis shows the retarded time minus the time at the first formation of apparent horizon, $t_\mathrm{AH}$. 
    }
    \label{fig_gw20}
\end{figure}

\begin{figure}
    \centering
    \includegraphics[width=0.98\columnwidth]{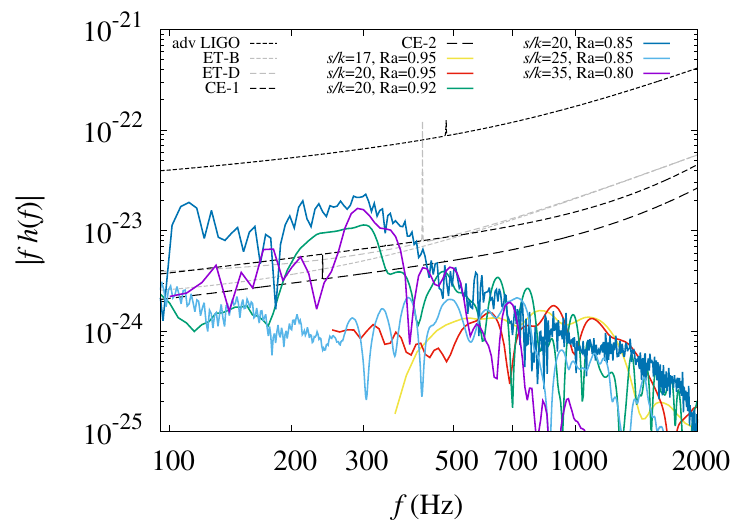}
    \vspace{-2mm}
    \caption{The effective amplitude $|f h(f)|$ as a function of frequency $f$ at a hypothetical distance to the source of 0.3\,Gpc for models with $(s/k, R_\mathrm{p}/R_\mathrm{e})=(17, 0.95)$, $(20, 0.95)$, $(20, 0.92)$, $(20, 0.85)$, $(20, 0.85)$, and ($35, 0.80)$. The designed sensitivities of the advanced LIGO, the ET, and the CE are also plotted.     }
    \label{fig_spec2d}
\end{figure}
\begin{figure}
    \centering
    \includegraphics[width=0.98\columnwidth]{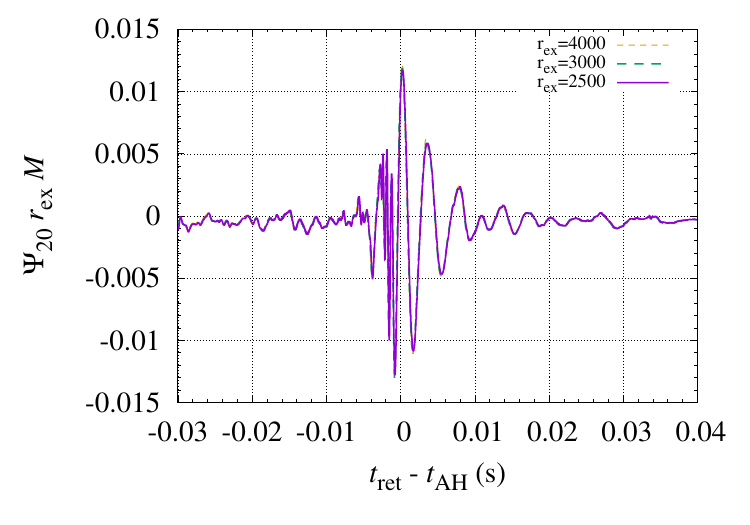}\\
    \vspace{-4mm}
    \includegraphics[width=0.98\columnwidth]{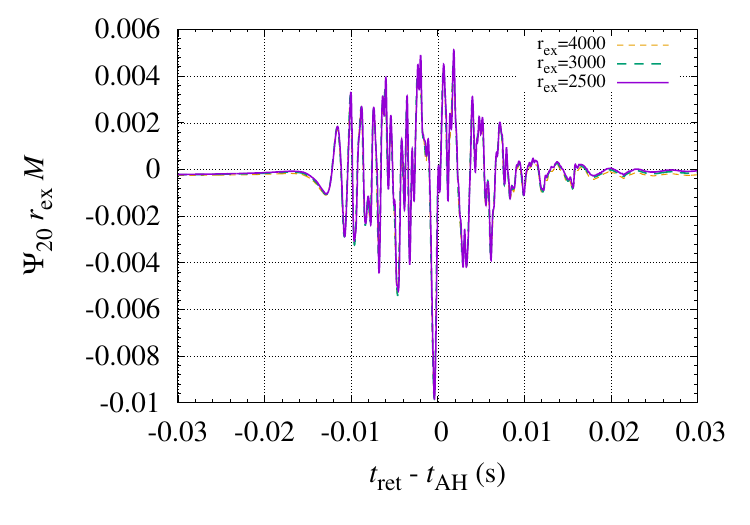}
    \vspace{-6mm}
    \caption{The same as Fig.~\ref{fig_gw20} but for model S20 with $R_\mathrm{p}/R_\mathrm{e}=0.85$ (upper panel) and 0.92 (lower panel).   }
    \label{fig_gw20.2}
\end{figure}

As summarized above, quasi-normal modes of black holes are excited at formation or during the early evolution of black holes, and, associated with this, ringdown gravitational waves are emitted. In addition, a dynamically oscillating disk and a dense ellipsoid can emit gravitational waves (see e.g.~\cite{Sykes:2026vmz}). The waveforms depend on which components are dominant. We find that these components predominantly generate high-frequency bursts in $\Psi_{20}$, and that these bursts are emitted only around the formation time of black holes (apparent horizons), denoted by $t_\mathrm{AH}$. Thus, in this subsection, we focus on $\Psi_{20}$ only around $t \sim t_\mathrm{AH}$. We note that low-frequency gravitational waves are emitted by the infall motion of matter and anisotropic outflow motion, but we do not focus on them because the frequency is likely to be less than 1\,Hz~\cite{2007ApJ...665L..43S}.  

Figure~\ref{fig_gw20} shows $\Psi_{20} r_\mathrm{ex}M$ as a function of retarded time minus $t_\mathrm{AH}$ for models S17 and S20 with $R_\mathrm{p}/R_\mathrm{e}=0.95$ and for model S35 with $R_\mathrm{p}/R_\mathrm{e}=0.80$. Here, the retarded time is defined by
\begin{equation}
t_\mathrm{ret}=t - r_\mathrm{ex}- 2M_\mathrm{ex} \ln (r_\mathrm{ex}/2M_\mathrm{ex}-1),
\end{equation}
and we choose $M_\mathrm{ex} (< M)$ so that waveforms at different extraction radii approximately align. 
For these models, a black hole forms without a long-lived dense spheroid, so the black hole mass at formation is relatively small, less than 10\% of the entire core mass, $M$. The disk fraction at black hole formation is also relatively small. Nevertheless, before black hole formation ($t_\mathrm{ret} < t_\mathrm{AH}$), gravitational waves with short wavelengths are emitted by the matter motion.
For these cases, a rotating neutron star with a maximum density exceeding the nuclear saturation density forms within $\alt 10\,\mathrm{ms}$ (see Fig.~\ref{fig_rhomax_ah}), and gravitational waves are emitted during its oscillation.
On the other hand, after black hole formation ($t_\mathrm{ret} > t_\mathrm{AH}$), a ringdown oscillation, likely associated with the black holes' quasi-normal modes, is observed. Notably, the characteristic frequencies of the oscillations for both $t_\mathrm{ret} < t_\mathrm{AH}$ and $t_\mathrm{ret} > t_\mathrm{AH}$ are quite high, $f_\mathrm{chr}\sim 0.5$--$1$\,kHz for models S17 and S20. This implies that $(2\pi f_\mathrm{chr} M)^{-2}$ is of order $10^{-2}$, and hence, the order of $h_{20}$ with $\Psi_{20}r_\mathrm{ex} M\sim 10^{-2}$ is
\begin{equation}
h_{20} \sim 10^{-24} \left({M \over 200M_\odot}\right)
\left({D \over 1\,\mathrm{Gpc}}\right)^{-1}. \label{eq28}
\end{equation}
For these models, the ringdown oscillation frequency is $\sim 1$\,kHz. Equation~\eqref{eq20} shows that the corresponding black hole mass is $\sim 14M_\odot$ assuming a dimensionless spin of 0.90--0.95. This estimated black hole mass is consistent with the black hole mass during the early evolution stage (cf. Fig.~\ref{fig_bh_evo}). 

For the S35 model, the frequency is below 1\,kHz ($\sim 300$\,Hz), but $M$ is large, so the factor $(2 \pi f_\mathrm{chr} M)^{-2} \sim 10^{-2}$ is small, and the order of magnitude of $h_{20}$ is given by Eq.~\eqref{eq28}. Since the mass is larger than in the previous two models, the amplitude can be effectively larger (cf.~Fig.~\ref{fig_spec2d}). 

All these features are more clearly visible in the Fourier spectrum in Fig.~\ref{fig_spec2d}. We indeed find the peak frequencies for models $(s/k, R_\mathrm{p}/R_\mathrm{e})=(17, 0.95)$ and $(20, 0.95)$ around 0.5--1\,kHz, and at $\sim 300$\,Hz for $(35, 0.80)$. Since the frequency is rather high and the amplitude is rather low for the (17, 0.95) and (20, 0.95) models, this type of gravitational wave is unlikely to be a promising source for ground-based detectors unless the collapse of rotating very-massive stars occurs in a nearby universe (at a distance of less than 100\,Mpc). The reason the frequency is high is that gravitational waves are emitted in the early stage of black hole evolution, when the black hole mass is rather small (less than 10\% of the core mass $M$: see Fig.~\ref{fig_bh_evo}) and the resulting disks around the black hole are compact. 

Figure~\ref{fig_gw20.2} is the same as Fig.~\ref{fig_gw20} but for model S20 with $R_\mathrm{p}/R_\mathrm{e}=0.85$ and $0.92$ (see also Fig.~\ref{fig_spec2d} for the effective amplitude of these models). For the $0.85$ model, the characteristic frequency, $f_\mathrm{chr} \sim 300$\,Hz, is much lower than for models S17 and S20 with $R_\mathrm{p}/R_\mathrm{e}=0.95$. This is because the black hole forms with a relatively large initial mass. For this model, a dense spheroid forms and survives for $\approx 80$\,ms before the black hole forms (see Fig.~\ref{fig_rhomax_ah}). As a result, the black hole mass in the early growth stage is relatively high, reducing the characteristic frequency. For this case, $(2\pi f_\mathrm{chr} M)^{-2} \sim 0.1$ with the characteristic frequency $\sim 300$\,Hz, and hence the possibility of detecting these gravitational waves is higher, although the effective amplitude is still not very high, $\sim 10^{-23}$, at a distance of $D=0.3$\,Gpc. We note that gravitational waves are excited even for $t_\mathrm{ret} < t_\mathrm{AH}$ for this model. These are emitted by the oscillating spheroid. It is found that the amplitude of this part is more than one order of magnitude smaller than that in the ringdown part.

For model S20 with $R_\mathrm{p}/R_\mathrm{e}=0.92$, the waveform feature is different from those discussed above; a ringdown waveform is not clearly seen, and gravitational waves associated with matter motion appear to be comparable to those by ringdown. This type of irregular waveforms is often found in the case that a dense spheroid is formed before the formation of the black hole, but the lifetime of the spheroid is relatively short, including the models with $(s/k, R_\mathrm{p}/R_\mathrm{e})=(17, 0.90)$ and $(20, 0.90)$. For this class of the waveforms, $\Psi_{20} r_\mathrm{ex} M \alt 0.01$, the frequency is several 100\,Hz--1\,kHz, and $(2\pi f_\mathrm{chr} M)^{-2} \alt 0.1$; because of the presence of many wave cycles with $f \sim 300$\,Hz, the effective amplitude is as large as that for the (20, 0.85) model. 

In any case, the amplitude of gravitational waves is lower than expected for $\epsilon \sim 1$ in Eq.~\eqref{eq19}; $\epsilon$ is of order $0.1$ for the axisymmetric oscillation. That is, only a fraction of the disk matter contributes to the emission of gravitational waves. Therefore, this type of gravitational wave is also unlikely to be a promising source for ground-based detectors unless the event occurs in a nearby universe. 

In summary, axisymmetric gravitational waves can reach fairly high amplitudes, around $\sim 10^{-23}$, at a hypothetical distance of $D\alt 1$\,Gpc for specific black hole formation scenarios, e.g., high-mass, rapidly rotating core collapse or collapses that produce a dense spheroid prior to black hole formation. Nevertheless, their characteristic frequency is quite high, ranging from several 100\,Hz--1\,kHz, even for very-massive stellar cores with $M=200$--$1100M_\odot$. Therefore, these gravitational waves are unlikely to be promising sources for the ET and the CE unless the event occurs at an exceptionally close distance. It is worth noting that, assuming the approximate scaling relations $f_\mathrm{chr} \propto M^{-1}$ and $h_{20} \propto M/D$ hold, the axisymmetric mode from collapse of very-massive, rapidly rotating stellar cores with $M \sim 3000$--$10^4M_\odot$ could be interesting sources for these detectors. A more systematic study on axisymmetric collapse is left for future work. 

\subsubsection{Waveforms for non-axisymmetric instabilities}

\begin{figure*}[p]
    \centering
    \includegraphics[width=0.91\columnwidth]{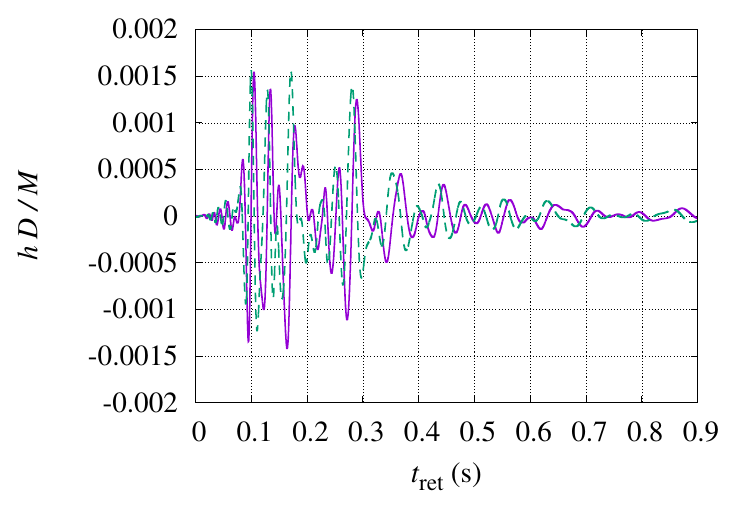}
    \includegraphics[width=0.91\columnwidth]{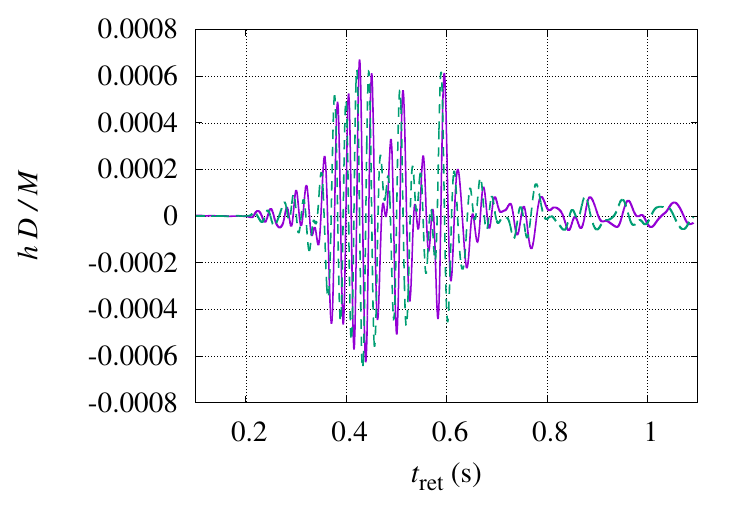} \\
    \vspace{-4mm}
    \includegraphics[width=0.91\columnwidth]{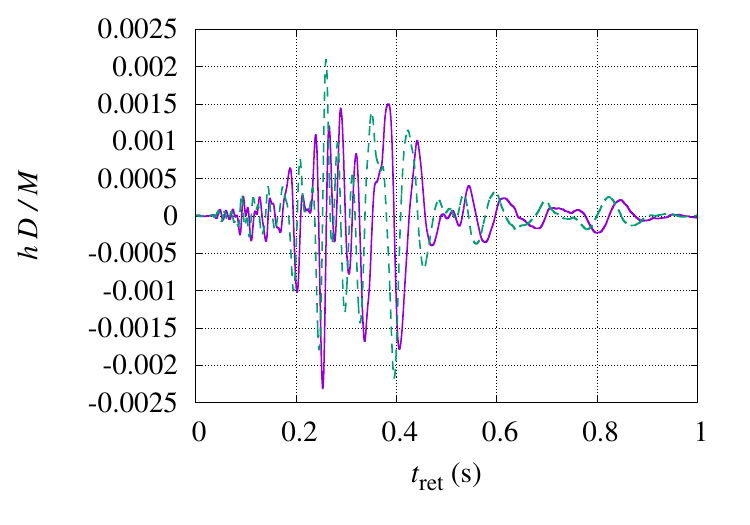}
    \includegraphics[width=0.91\columnwidth]{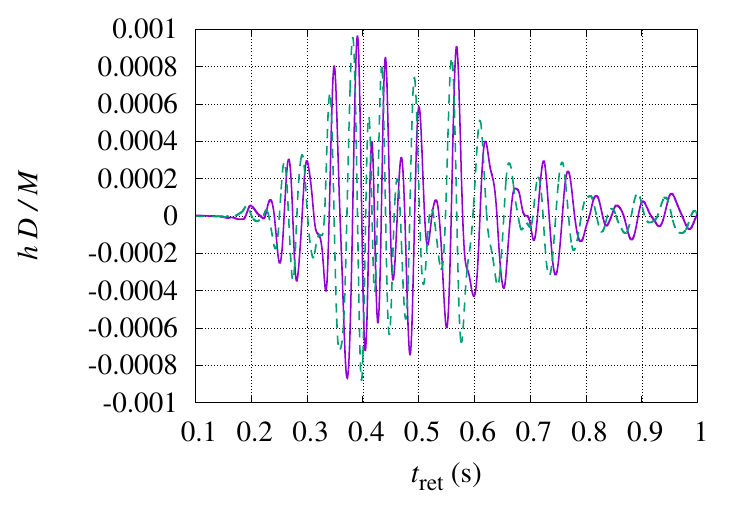}   \\
    \vspace{-4mm}
    \includegraphics[width=0.91\columnwidth]{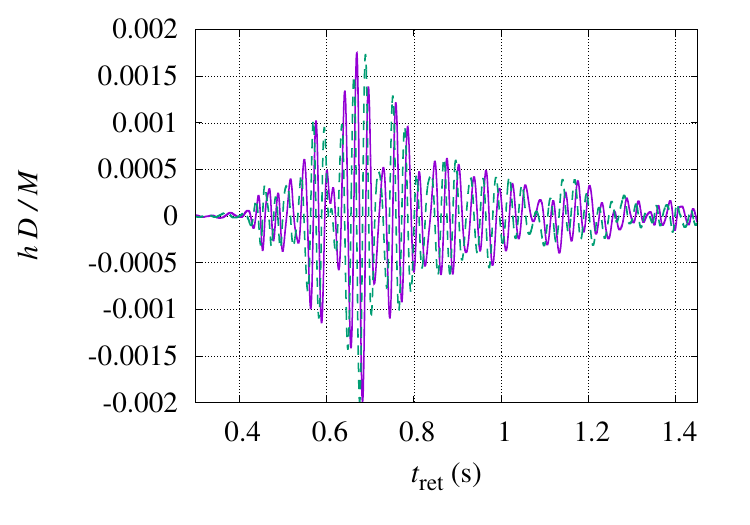}    \includegraphics[width=0.91\columnwidth]{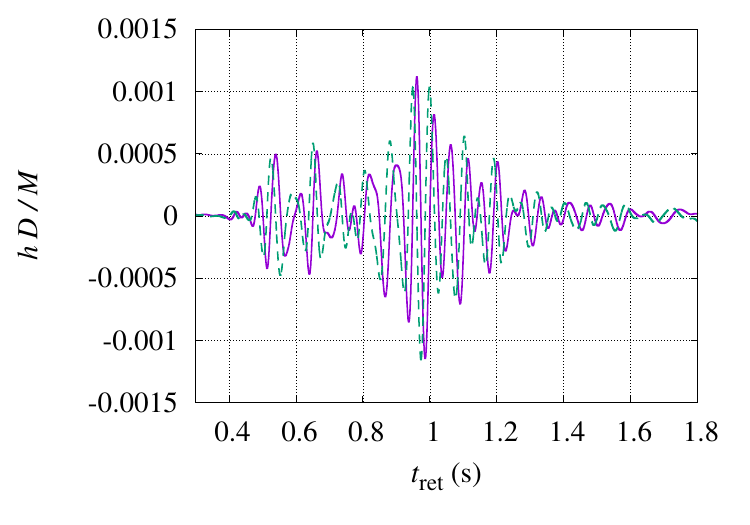}\\
        \vspace{-4mm}
    \includegraphics[width=0.91\columnwidth]{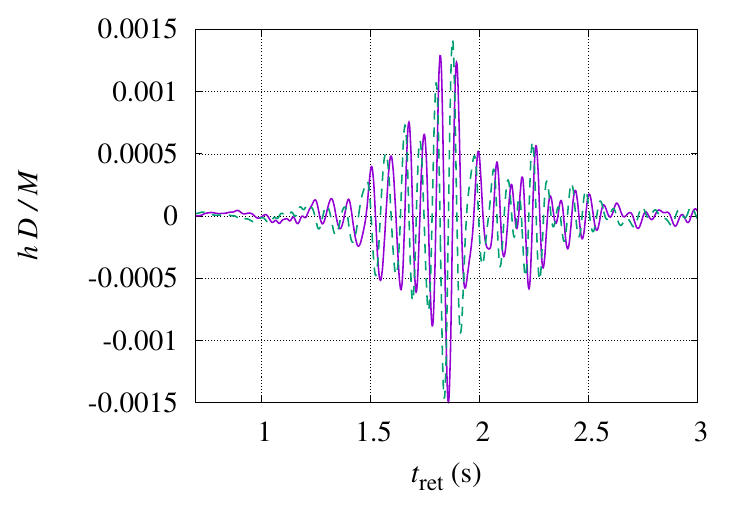}    \includegraphics[width=0.91\columnwidth]{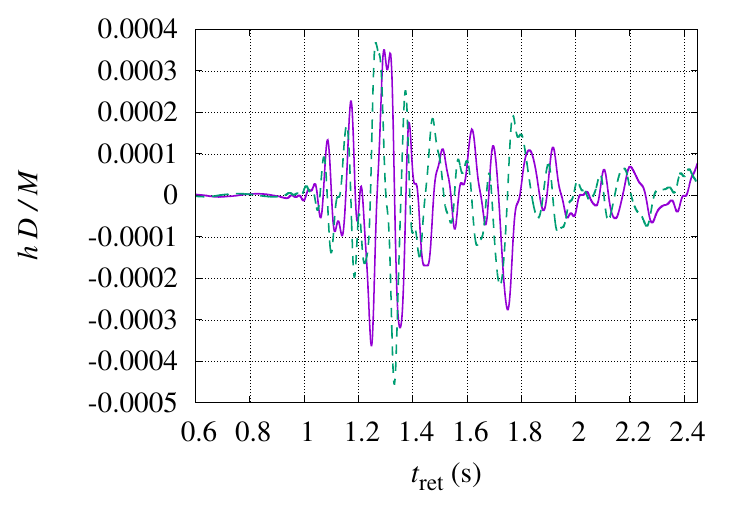}
    \vspace{-4mm}
    \caption{Gravitational waveforms of the $l=m=2$ mode from non-axisymmetrically unstable disks in units of $M/D(=GM/c^2D)$ where $D$ denotes the distance to the source. Top row panels: for models S17 with $R_\mathrm{p}/R_\mathrm{e}=0.90$ (left) and 0.95 (right). Second row panels: for models S20 with $R_\mathrm{p}/R_\mathrm{e}=0.85$ (left) and 0.92 (right). Third row panels: models S25 with $R_\mathrm{p}/R_\mathrm{e}=0.85$ (left) and 0.90 (right). Bottom row panels: models S35 with $R_\mathrm{p}/R_\mathrm{e}=0.75$ (left) and $0.90$ (right). The solid and dashed curves denote the plus and cross modes, respectively. We assume the detection of gravitational waves along the $z$-axis (most optimistic direction). For these plots, $t=0$ implies the start time of the 3D simulations. The extraction radius for these waveforms is $r_\mathrm{ex}=2\times 10^4M_\odot$. 
    }
    \label{fig_gw}
\end{figure*}

In Fig.~\ref{fig_gw}, we display gravitational waveforms from 3D numerical simulations for 8 models with $(s/k, R_\mathrm{p}/R_\mathrm{e})=(17, 0.90)$, $(17, 0.95)$, $(20, 0.85)$, $(20, 0.92)$, $(25, 0.85)$, $(25, 0.90)$, $(35, 0.75)$, and $(35, 0.90)$. As mentioned in Sec.~\ref{sec4.4}, all simulations start after black hole formation, i.e., at $t>t_\mathrm{AH}$. Figure~\ref{fig_gw} shows that the waveforms universally consist of a couple of burst waves excited when a non-axisymmetric instability develops, followed by quasi-periodic waves of smaller amplitude, although the detailed waveforms depend on the black hole and disk formation process. The peak gravitational-wave amplitude occurs when the non-axisymmetric instability saturates.

As indicated in Eq.~\eqref{eq19}, a higher amplitude is achieved for more massive and more compact disks. To this end, more rapidly spinning progenitor cores are preferable, and thus, for a given value of $R_\mathrm{p}/R_\mathrm{e}$, less massive cores have an advantage in enhancing $h D/M(=h (c^2D/GM))$. We also observe that the waveform varies depending on when the 3D simulations begin, especially for rapidly rotating models where instability can develop shortly after black hole formation due to early massive disk formation (see Fig.~\ref{fig_disk_mass_ratio}). Nonetheless, the characteristic frequencies and the spectrum peak amplitude do not depend strongly on the simulation start time.

\begin{figure*}[b]
    \centering
    \vspace{-0.5cm}
    \hspace{-0.5cm}
    \includegraphics[width=1.18\columnwidth]{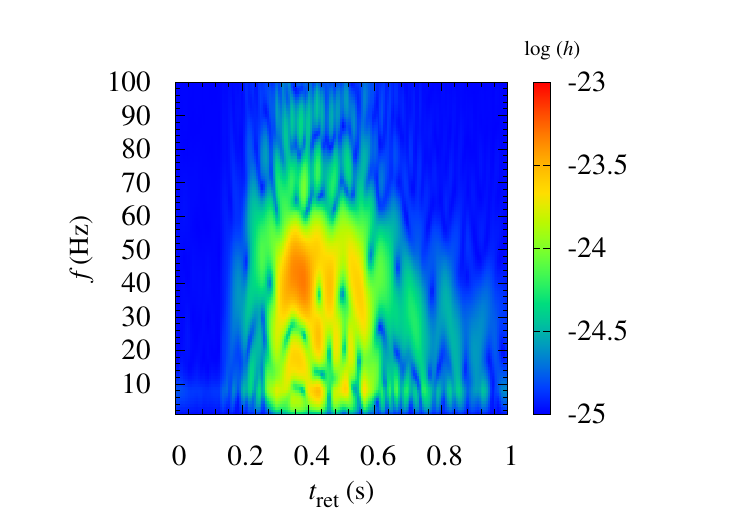}
    \hspace{-2.4cm}
    \includegraphics[width=1.18\columnwidth]{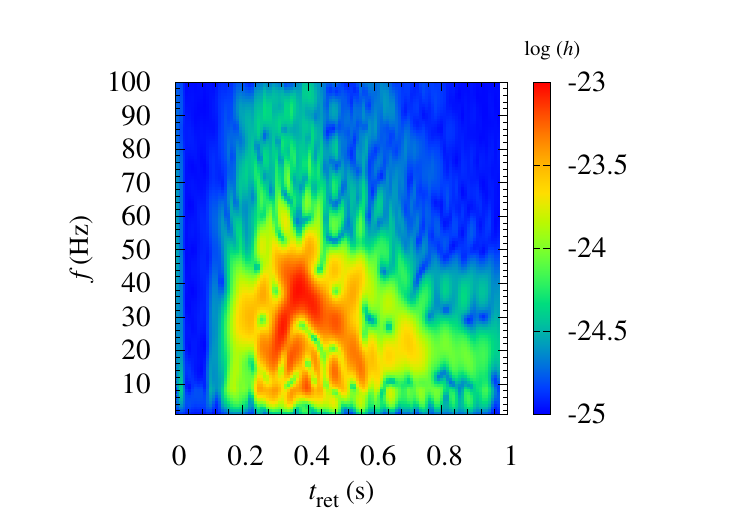} \\
    \vspace{-6mm}
    \hspace{-0.5cm}
    \includegraphics[width=1.18\columnwidth]{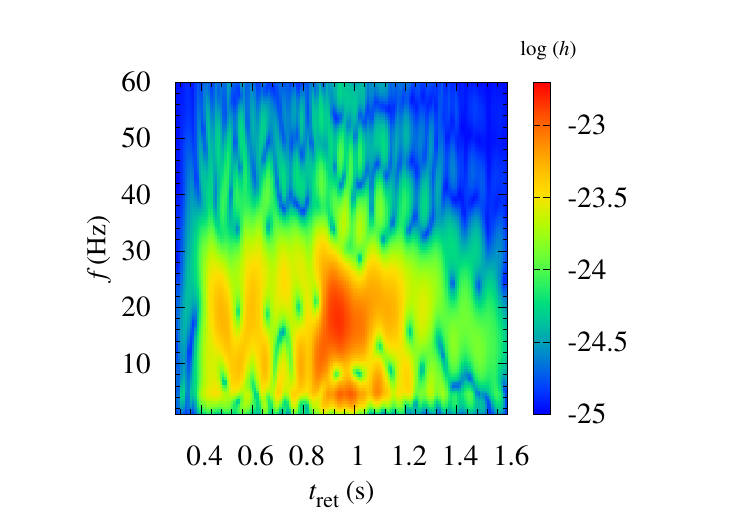}
    \hspace{-2.4cm}
    \includegraphics[width=1.18\columnwidth]{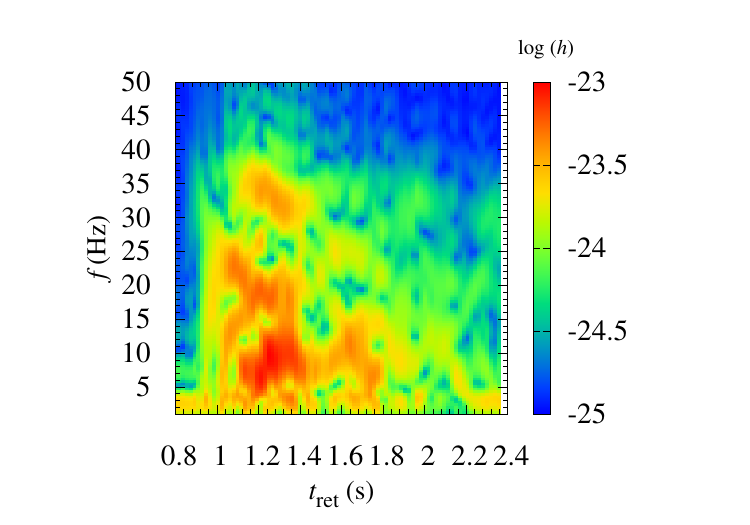}  
    \vspace{-2mm}
    \caption{Spectrograms of gravitational waves for the waveforms displayed in the right side of Fig.~\ref{fig_gw}, i.e., for $(s/k, R_\mathrm{p}/R_\mathrm{e})=(17, 0.95)$ (top left), $(20,0.92)$ (top right), $(25,0.90)$ (bottom left), and $(35, 0.90)$ (bottom right). The color indicates the effective amplitude of gravitational waves at the hypothetical distance to the source of 3\,Gpc for all the plots.
    }
    \label{fig_spg}
\end{figure*}
\begin{figure*}[th]
    \centering
    \includegraphics[width=0.99\columnwidth]{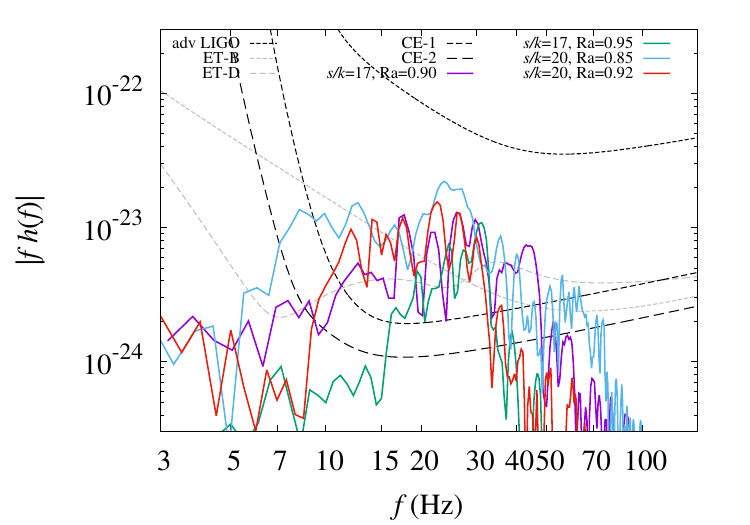}
    \includegraphics[width=0.99\columnwidth]{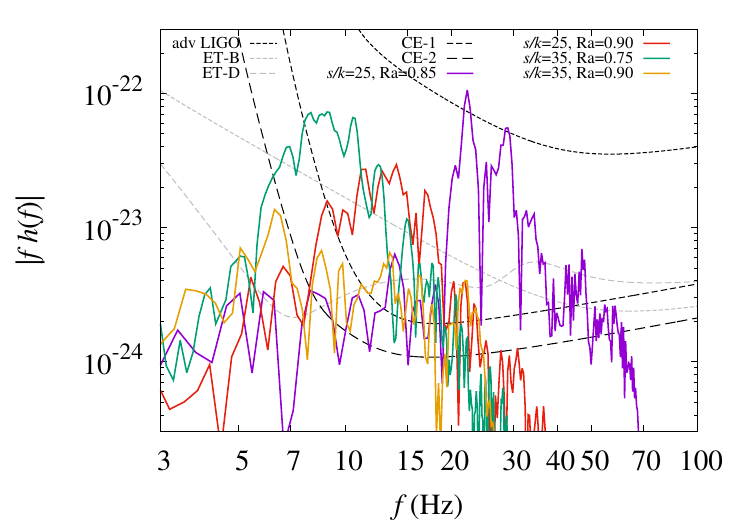}    
    \vspace{-2mm}
    \caption{Spectrum (effective amplitude) of gravitational waves in terms of $|h(f)f|$ from non-axisymmetrically unstable systems for $(s/k, R_\mathrm{p}/R_\mathrm{e})=(17, 0.90)$, $(17,0.95)$, $(20, 0.85)$, and $(20, 0.92)$  (left) and $(25, 0.85)$, $(25, 0.90)$, $(35, 0.75)$, and $(35, 0.90)$ (right). We assume that the hypothetical distance to the source is 3\,Gpc (cosmological redshift, $z_\mathrm{cos}\approx 0.51$) and the observation is done along the $z$-axis (most optimistic direction). The designed sensitivities of advanced LIGO, ET, and CE ($(S_\mathrm{n}(f)f)^{1/2}$ with $S_\mathrm{n}(f)$ being the one-sided noise spectrum density in this figure) are also plotted. The cosmological redshift effect is taken into account for the spectrum; i.e., the spectrum is plotted for $f=f_0/(1+z_\mathrm{cos})$ with $f_0$ the frequency in the source frame.
    }
    \label{fig_spec}
\end{figure*}

A clear difference is observed in the gravitational waveforms between low-mass (S17 and S20) and high-mass (S25 and S35) models. For the low-mass cases, there are a couple of high-amplitude waves with comparable peak amplitudes, whereas for the high-mass cases, there is only a single-peak wave. Our interpretation of this difference is summarized as follows: Figure~\ref{fig_bh_evo} shows that the black hole mass evolution timescale is always several 100\,ms, irrespective of the total mass of the system. This implies that, when time is normalized by $GM/c^3$, the timescale for the evolution of the black hole mass is longer for smaller values of $M$, reflecting the density profile of the progenitor cores (i.e., lower-mass progenitor cores have more extended density profiles; cf. Fig.~\ref{fig0}). Therefore, for the smaller-mass models, the evolution of the disk due to matter infall proceeds more slowly (in terms of $GM/c^3$). As a result, a large-mass disk forms at a relatively late stage of the evolution of the system for the lower-mass models, leading to the emission of high-amplitude gravitational waves over a broad period.

For models with $(s/k, R_\mathrm{p}/R_\mathrm{e})=(17, 0.90)$ (top left) and $(20, 0.85)$ (second-row left), gravitational waves are excited at $t_\mathrm{ret} \sim 0$. In these cases, a dense spheroid forms with a lifetime longer than 50 ms before black hole formation. As a result, the disk is already massive at the start of the 3D simulation and primed for potential instability (see Fig.~\ref{fig_disk_mass_ratio}). This suggests that the dense spheroid could be susceptible to non-axisymmetric deformation, although this aspect is not explored in our current study. For other models, the disk generally becomes significantly unstable as its mass grows due to infall. 

The peak amplitude of gravitational waves is typically of order $10^{-3}$ of $GM/(c^2D)$, i.e., 
\begin{align}
h_\mathrm{peak} &\sim 10^{-3} {GM \over c^2D} \epsilon_{-3}\nonumber \\
&\approx 1.4 \times 10^{-23}\left({M \over 300M_\odot}\right)
\left({D \over 1\,\mathrm{Gpc}}\right)^{-1} \epsilon_{-3},
\end{align}
where $\epsilon_{-3}$ is a factor of order unity. 
This amplitude is accessible to the ET and CE, especially in the high-mass disk formation. The gravitational waveform exhibits a quasi-periodic pattern with $f \sim 40 (M_\mathrm{BH}/200M_\odot)^{-1}$\,Hz (see the spectrum in Fig.~\ref{fig_spec}). This aligns roughly with Eq.~\eqref{eq13} for $\varpi \sim 4GM_\mathrm{BH}/c^2$, and is much lower than the axisymmetric gravitational wave frequencies discussed earlier. Although the black hole's mass exceeds that of stellar-mass black holes of $O(10M_\odot)$, the characteristic frequency still falls within the sensitive range of ground-based detectors, particularly the ET and CE. This is due to the high dimensionless spin of the resulting black hole, which facilitates the formation of a compact disk and increases the typical frequency.

Figure~\ref{fig_spg} presents the spectrograms of the waveforms shown on the right side of Fig.~\ref{fig_gw}. These spectrograms reveal that the main frequencies of the gravitational waves are clearly visible and may change over time or manifest as multiple characteristic frequencies. For models S17 with $R_\mathrm{p}/R_\mathrm{e}=0.95$ (top left) and S25 with $R_\mathrm{p}/R_\mathrm{e}=0.90$ (bottom left), the dominant frequencies remain approximately 40 Hz and 20 Hz, respectively. In contrast, model S20 with $R_\mathrm{p}/R_\mathrm{e}=0.92$ shows a rising dominant frequency from 20 Hz to 40 Hz between $t_\mathrm{ret} \sim 0.25$ s and $\sim 0.35$ s, then it decreases again. Meanwhile, model S35 with $R_\mathrm{p}/R_\mathrm{e}=0.90$ exhibits three characteristic frequencies around 10, 20, and 30 Hz during $t_\mathrm{ret} \sim 1.2$--1.4 s. These variations are interpreted as reflecting the different formation processes of the disks around the black hole. Notably, most of the spectrogram shapes shown in this paper differ qualitatively from those in Ref.~\cite{Shibata:2021sau}, which examined gravitational waves emitted by unstable tori initially in equilibrium around a black hole. That study found spectrograms mainly feature a spot in a narrow frequency range. In that straightforward scenario, the key frequency depends solely on the unstable mode of the tori. Conversely, in the current context, the characteristic frequencies are influenced by the evolving disks caused by matter infall from outside, allowing them to change over time.

The Fourier spectra of gravitational waves are shown in Fig.~\ref{fig_spec} for a hypothetical source distance of 3\,Gpc (cosmological redshift, $\approx 0.51$). We plot the dimensionless spectrum (effective amplitude) $|fh(f)|$ and compare its amplitude with the design sensitivities of advanced LIGO, the ET, and the CE in the figure. 
It is found that the spectrum shape is qualitatively universal: It peaks at a characteristic frequency associated with the non-axisymmetrically unstable modes of the disk, and the amplitude declines toward higher frequencies, eventually reaching a sharp cutoff determined by the maximum orbital frequency of the disk around the formed black hole. We note that the low-frequency side of the spectrum, far below the peak, is not reliable due to the limited computational time in the 3D simulations, typically $\alt 2$\,s.
For lower-mass models (S17 and S20), the spectrum has a broader peak (except for model S17 with $R_\mathrm{p}/R_\mathrm{e}=0.95$). This reflects the properties of the waveforms: For these models, the waveforms consist of multiple peaks at different times, whereas for higher-mass models (S25 and S35), the maximum amplitude is achieved only once (see Fig.~\ref{fig_gw}).

At the hypothetical distance of $D=1$\,Gpc, the peak amplitude is appreciably higher than the designed noise levels of the ET and the CE. Even for $D=$a few Gpc, the signal-to-noise ratio (SNR), defined by (e.g., Ref.~\cite{Read:2013zra}) 
\begin{equation}
\mathrm{SNR}=\sqrt{4 \int df {|h(f)|^2 \over S_\mathrm{n}(f)}}
\end{equation}
with $S_\mathrm{n}(f)$, the one-sided noise spectrum density, can exceed 10 for some models, particularly large-mass, rapidly rotating models (see Table~\ref{tab1}). This indicates that the collapse of rotating, very-massive stellar cores is a candidate source for these future detectors, even when the event occurs at cosmological distances of $\agt 1$\,Gpc. 

Higher-mass models have a lower characteristic frequency. Nonetheless, even in models S35, the peak frequency remains between 10 and 20 Hz, aligning with the sensitive range of ET and CE. For models S17, S20, and S25, the peak frequencies are between 20 and 70 Hz, placing the gravitational waves within the most sensitive band of these detectors.

The gravitational-wave frequencies reported in this paper are by a factor of 2--3 higher than our prediction~\cite{Shibata:2025lde}. This discrepancy arises because the compactness of the disk orbits around the rapidly spinning black hole exceeds the predicted values. As a result, the numerical results in this paper are more encouraging for detecting gravitational waves from the core collapse of very-massive stars. 

\section{Summary and discussion}\label{sec5}

By performing both axisymmetric and 3D numerical relativity simulations, we studied the fate of the collapse of rotating very-massive stellar cores with mass $1100M_\odot$, varying the degree of rotation over a wide range. 
The axisymmetric simulations reveal two classes of outcomes in this problem. One is the formation of a black hole, and the other is an oscillating spheroid with a maximum density less than $\sim 10^9\,\mathrm{g/cm^3}$. The latter typically occurs in rapidly rotating models, and the spheroid forms via centrifugal bounce. We found that centrifugal bounce is more likely in lower-mass very-massive stars, and for high-mass very-massive stars with core masses larger than $\sim 2000M_\odot$, this outcome is likely absent if we assume rigid rotation of the progenitor cores. Our current study does not include details of neutrino cooling and nuclear burning, so the evolution of the oscillating spheroid could be modified in a more detailed study, which is one of our future topics. 

In models with moderate or rapid rotation, the resulting black hole is encircled by a substantial disk that may be prone to non-axisymmetric instabilities. The 3D simulations confirmed the presence of these models. 
Our research indicates that the process of black hole formation is influenced by the degree of rotation. When rotation is relatively rapid, a dense spheroid temporarily appears just prior to black hole formation, with a lifetime less than 100\,ms. If the dense spheroid does not form, the initial black hole mass remains very small, typically under 10\% of the total core mass. Conversely, when the dense spheroid does form, the initial black hole tends to be substantially larger. 

Our detailed analysis of gravitational waves revealed that even without non-axisymmetric deformation, axisymmetric modes with appreciable amplitudes are emitted. This mode exhibits a high amplitude around the black hole formation time, peaking at $\sim 10^{-23}$ at a hypothetical distance of 1 Gpc. Despite the system's high mass, the axisymmetric mode remains at a high frequency of roughly 300–1000 Hz because the mass in the central region, including the black hole, is much less than the entire core mass at the moment of black hole formation. 

By contrast, gravitational waves from disks undergoing non-axisymmetric deformation have low frequencies, typically $\sim 10$--50\,Hz. The amplitude can reach $\sim 10^{-22}$ at a hypothetical distance of 1\,Gpc, i.e., by one order of magnitude higher than those of axisymmetric modes. This type of gravitational wave is universally composed of high-amplitude burst gravitational waves in the first several cycles, followed by subsequent quasi-periodic waves of lower amplitude. Since the expected amplitude is high and the frequencies are within the sensitive band, this type of gravitational wave can be a promising source for the ET and the CE, even if the core collapse of very-massive stars occurs at $z \sim 0.5$ ($z$ here denotes the cosmological redshift). 

The massive disk is not only a source of gravitational waves but also a possible engine for energetic electromagnetic transients. Thus, the collapse of rotating very-massive stars can be a multi-messenger transient source. In the presence of magnetic fields in the disk, it is natural to suppose that turbulence develops via the magnetorotational instability~\cite{Balbus:1998ja} and that the disk behaves as a viscous fluid. Since viscous heating should be most efficient near the black hole, convective motion associated with the entropy gradient is likely to develop in the disk, leading to a mass outflow from the innermost region of the disk with the typical velocity of $v_\mathrm{out} \sim 0.05c$ (see, e.g., Refs.~\cite{Fernandez2013a, Just2015a, Fujibayashi2020a, Fujibayashi2020b, Fujibayashi:2023oyt} for similar phenomena). Since the outflow mass $M_\mathrm{out}$ can be large, the kinetic energy of the outflow could be huge as
\begin{eqnarray}
E_\mathrm{out}&\approx& {1 \over 2}M_\mathrm{out}v^2_\mathrm{out} \nonumber \\
&\approx& 1.1 \times 10^{53}\,\mathrm{erg}\left({M_\mathrm{out} \over 50M_\odot}\right)
\left({v_\mathrm{out} \over 0.05c}\right)^2.
\end{eqnarray}
During the evolution of very-massive stars, about half of the initial stellar mass forms an extended envelope composed of hydrogen and helium, with a radius of order $R_\mathrm{env} \sim 10^{14}$--$10^{15}$\,cm, around the central core (e.g., Ref.~\cite{Takahashi2018}). Under this assumption for the envelope mass, say $M_\mathrm{env}\sim 300M_\odot$ for models S20, the resulting envelope velocity after injection of the disk outflow would be $v \sim v_\mathrm{out} \sqrt{M_\mathrm{out}/(M_\mathrm{out}+M_\mathrm{env})}$. Since a substantial fraction of the outflow's kinetic energy is likely to be dissipated within it, such an envelope should shine similarly to type II supernovae, with the characteristic time $\tau_\mathrm{chr}$, defined by $\tau_\mathrm{exp}=\tau_\mathrm{diff}$, where $\tau_\mathrm{exp}$ and $\tau_\mathrm{diff}$ are the expansion and diffusion timescales, respectively~\cite{Arnett1982} as
\begin{align}
\tau_\mathrm{chr} \sim \sqrt{{3M_\mathrm{tot} \kappa \over 4\pi c v}} 
&\approx 3 \times 10^7\,\mathrm{s} ~\kappa_{0.1}^{1/2}
\left({M_\mathrm{tot} \over 350M_\odot}\right)^{3/4}
\nonumber \\
&\times 
\left({M_\mathrm{out} \over 50M_\odot}\right)^{-1/4}
\left({v_\mathrm{out} \over 0.05c}\right)^{-1/2}, 
\end{align}
where $M_\mathrm{tot}=M_\mathrm{env}+M_\mathrm{out}$ and $\kappa$ denotes the typical opacity of the envelope with $\kappa_{0.1}=\kappa/0.1\,\mathrm{cm^2\,g^{-1}}$. Since $\tau_\mathrm{chr} v$ is much larger than the radius of the envelope supposed, the adiabatic cooling of the expanding envelope would reduce the internal energy of the envelope before the peak time, but still the luminosity at $t=\tau_\mathrm{chr}$ is comparable to or larger than that of typical type II supernovae as
\begin{eqnarray}
L_\mathrm{chr} &\sim& {E_\mathrm{out} \over 2 \tau_\mathrm{chr}}\times {R_\mathrm{env} \over  \tau_\mathrm{chr} v} \nonumber \\
&\approx& 5 \times 10^{43}\,\mathrm{erg/s}~~\kappa_{0.1}^{-1}
\left({R_\mathrm{env} \over 5 \times 10^{14}\,\mathrm{cm}}\right)
\left({v_\mathrm{out} \over 0.05c}\right)^{2} \nonumber \\
&&~~~~~~~\times
\left({M_\mathrm{out} \over 50M_\odot}\right)
\left({M_\mathrm{tot} \over 350M_\odot}\right)^{-1},
\label{eq17}
\end{eqnarray}
where we assumed that half of $E_\mathrm{out}$ is converted into internal energy due to the dissipation of kinetic energy. 
This estimate is consistent with that in Ref.~\cite{Uchida:2018ago}. Thus, the explosion is likely to be observed as a year-long transient as bright as or brighter than typical supernovae. 
Such a disk outflow may also synthesize heavy elements if the disk density is high enough to be neutron-rich~\cite{2025arXiv250315729A}. 

We note that radioactive heating, e.g., from $^{56}$Ni that could be produced in the disk wind, may contribute to the heating of the ejecta~\cite{Fujibayashi:2026}. However, the total energy available for this is likely to be more than an order of magnitude smaller than $E_\mathrm{out}$ because only a fraction of the outflow mass can be $^{56}$Ni, and the specific energy release from $^{56}$Ni decay is $\sim 10^{-4}c^2$, which is an order of magnitude smaller than the approximate specific kinetic energy, $v_\mathrm{out}^2/2c^2\sim 10^{-3}$. Thus, radioactive heating would be subdominant. It may become more important if the progenitor envelope is compact, for example, as a result of binary interaction.

To estimate its contribution on the characteristic diffusion time $\tau_\mathrm{chr}$, we compare the shock-deposited internal energy remaining in the ejecta, $E_\mathrm{int}(\tau_\mathrm{chr}) \approx (E_\mathrm{out}/2)(R_\mathrm{env}/v\tau_\mathrm{chr})$, with the radioactive energy supplied over an expansion timescale near $\tau_\mathrm{chr}$, $\Delta E = \dot{E}_\mathrm{Ni}(\tau_\mathrm{chr})\cdot\tau_\mathrm{chr}$.
Since $\tau_\mathrm{chr}$ for the considered parameter is much longer than the decay lifetime of $^{56}$Ni ($\approx 8.8$\,d), $\Delta E_\mathrm{chr} \approx 7.3\times10^{-5}M_\mathrm{Ni}c^2 e^{-\tau_\mathrm{chr}/\tau_\mathrm{Co}}\cdot (\tau_\mathrm{chr}/\tau_\mathrm{Co}$), where $\tau_\mathrm{Co}=111.3$\,d \citep{1994ApJS...92..527N}. Radioactive heating may affect the light curve if the ratio $\Delta E_\mathrm{chr}/E_\mathrm{int}(\tau_\mathrm{chr}) \approx 3\times10^{-4}(M_\mathrm{Ni}/M_\mathrm{tot})(c^2/v^2)(v\tau_\mathrm{chr}/R_\mathrm{env})(\tau_\mathrm{chr}/\tau_\mathrm{Co})e^{-\tau_\mathrm{chr}/\tau_\mathrm{Co}}$ becomes unity, or if the envelope radius satisfies
\begin{align}
R_\mathrm{env} &\lesssim 3\times10^{-4} v\tau_\mathrm{chr} \frac{M_\mathrm{Ni}}{M_\mathrm{tot}} \frac{c^2}{v^2}\frac{\tau_\mathrm{chr}}{\tau_\mathrm{Co}}e^{-\tau_\mathrm{chr}/\tau_\mathrm{Co}},\notag\\
&\approx 2\times10^{15}\,\mathrm{cm}\,\frac{M_\mathrm{Ni}}{M_\mathrm{tot}},
\end{align}
where the parameter values used in the above estimates are also used in the second expression. Thus, for $R_\mathrm{env}=5\times10^{14}$\,cm, the nickel mass comparable to the assumed outflow mass would be required. Therefore, the radioactive heating is expected to modify the light curve only if an exceptionally large fraction of the outflow is composed of $^{56}$Ni, or if the progenitor envelope is substantially compact.

After the development of the magnetohydrodynamical turbulence in the disk, the disk matter together with magnetic fluxes fall into the black hole, and subsequently, a black hole magnetosphere composed of a global poloidal magnetic field is likely to be formed~\cite{Christie2019dec, Hayashi:2021oxy}. Assuming that the equi-partition is established in the disk (see, e.g., Ref.~\cite{Shibata:2025gix}), the typical magnetic field strength determined from $B^2/(8\pi) \sim \eta \rho_\mathrm{max} c_\mathrm{s}^2$ would be
\begin{equation}
B \sim 10^{14}\,\mathrm{G} \left({\rho_\mathrm{max} \over 10^{10}\,\mathrm{g/cm^3}}\right)^{1/2}
    \left({c_\mathrm{s} \over 10^9\,\mathrm{cm/s}}\right),\label{eq18}
\end{equation}
where $\rho_\mathrm{max}$ is the maximum density, $c_\mathrm{s}$ is the typical sound velocity around the density maximum of the disk, and $\eta$ is a constant of order $10^{-2}$ for the turbulence associated with the magnetorotational instability: For Eq.~\eqref{eq18}, we set $\eta=0.04$. 
Assuming that a magnetosphere with a poloidal magnetic field of $B\sim 10^{14}$\,G is established, the Blandford-Znajek luminosity~\cite{Blandford1977} is estimated as
\begin{eqnarray}
    L_\mathrm{BZ}
    &\sim & 1 \times 10^{52}\,{\rm erg/s}
    \left({M_{\rm BH} \over 100M_\odot}\right)^2 
    \left({B_\mathrm{p} \over 10^{14}\,{\rm G}}\right)^2
    \left({\chi_\mathrm{BH} \over 0.9}\right)^2. \nonumber \\\label{eq10}
\end{eqnarray}
Thus, a strong energy injection is expected. We note that in the present scenario, the dimensionless spin of the black hole is likely to be quite large, which is conducive to larger energy injection. 

The total energy of the black hole available for the Blandford-Znajek mechanism is approximately written as~\cite{Shibata:2023tho}
\begin{equation}
E_\mathrm{BH} \approx 1\times 10^{55} \,\mathrm{erg}\,
\left({\chi_\mathrm{BH} \over 0.9}\right)^2 
\left({M_\mathrm{BH} \over 100M_\odot}\right). 
\label{eq7}
\end{equation}
Thus, the energy injection is likely to continue for the order of 
\begin{equation}
E_\mathrm{BH}/L_\mathrm{BZ} \sim 10^3\,\mathrm{s}
\left({M_\mathrm{BH} \over 100M_\odot}\right)^{-1}
\left({B_\mathrm{p} \over 10^{14}\,{\rm G}}\right)^{-2},
\end{equation}
which is likely to be shorter than $R_\mathrm{env}/c< R_\mathrm{env}/v$.  
If this energy injection successfully penetrates the envelope of a very-massive star, an ultra-long gamma-ray burst may be observed~\cite{2011ApJ...726..107S}. If the jet is choked, this energy injection, which can be much larger than that from the disk outflow, will heat the envelope, increasing the expansion velocity $v$ by an order of magnitude (i.e., $v> 0.1c$). As a result, the stellar explosion may be much brighter than estimated in Eq.~\eqref{eq17}, with $L_\mathrm{chr} > 10^{45}$\,erg/s. During the expansion of the ejected matter, interstellar matter is accumulated, forming shocks, and synchrotron radiation is likely to be induced. This component is likely to be very bright as well because of the fast motion of the ejecta. In future work, we plan to clarify the effect of jets in this scenario. It is also important to examine whether the high magnetic field strength shown in Eq.~\eqref{eq18} is actually achieved. 

In this work, we employ simplified equations of state to model the collapsing core. This approach is reasonable for studying the collapse stage but may not be appropriate for the evolution of spheroids formed by centrifugal bounces. In rapidly rotating core collapse of relatively low-mass very-massive stars, the centrifugal bounce occurs when the central density is $10^7$--$10^9\,\mathrm{g/cm^3}$. Because the lifetime of the resulting spheroid is much longer than the dynamical timescale in this case, the effects of nuclear burning and neutrino cooling may be crucial to its subsequent evolution. A simulation with more detailed microphysics is necessary to clarify the evolution of such spheroids. Since the dense spheroid with central density $10^{13}$--$10^{14}\,\mathrm{g/cm^3}$ has a short lifetime of 50--100\,ms, we expect that microphysical effects would not drastically modify the evolution in this case. Nevertheless, microphysical effects could determine the lifetime of the dense spheroids. 

In addition, both types of spheroids could be unstable to non-axisymmetric deformation, a phenomenon not studied in this paper. It should be noted that unstable dense spheroids may be a source of gravitational waves for the ET and the CE because the mass and equatorial radius are of order $M_\mathrm{sphe} \sim 10M_\odot$ and $R_\mathrm{sphe} \sim 100$\,km, and hence the expected gravitational-wave frequency is 
\begin{align}
f_\mathrm{sphe} &\approx {1 \over \pi} \sqrt{{GM_\mathrm{sphe} \over R_\mathrm{sphe}^3}} \nonumber\\
&\approx 367
\left({M_\mathrm{sphe} \over 10M_\odot}\right)^{1/2}\left({R_\mathrm{sphe} \over 100\,\mathrm{km}}\right)^{-3/2}
\,\mathrm{Hz}. 
\end{align}
Therefore, in the presence of a bar mode, the gravitational wave frequency would be several 100\,Hz. A more detailed study of the evolution of dense spheroids is left for future work. 

\acknowledgements

We thank the members of the Computational Relativistic Astrophysics division for the daily discussions. 
Numerical computation was performed on Sakura and Momiji clusters at the Max Planck Computing and Data Facility. This work was in part supported by Grant-in-Aid for Scientific Research (grant Nos.~23H04900, 26K00732) of Japanese MEXT/JSPS. Lam acknowledges support by NASA under award No. 80NSSC25K7213.

\bibliography{reference}

\end{document}